\documentclass[aps,reprint]{revtex4-2} 
\usepackage{graphicx,graphics} 
\usepackage{epstopdf} 
\usepackage{dcolumn}
\usepackage{amsmath,amssymb,amsfonts} 
\usepackage{latexsym,verbatim} 
\usepackage{bm} 
\usepackage{bbold} 
\usepackage{xcolor} 
\usepackage{ulem} 
\usepackage[breaklinks=false,colorlinks,citecolor=blue,linkcolor=blue,urlcolor=blue]{hyperref} 
\usepackage[abs]{overpic} 
\usepackage{textcomp} 
\usepackage{mathtools} 
\usepackage{braket} 
\usepackage{soul} 
\usepackage{lipsum}
\usepackage[export]{adjustbox}
\usepackage{gensymb}
\usepackage[T1]{fontenc}
\usepackage{multirow}

\begin{document}

\title{Engineering entanglement and transport in interacting quantum walks with tailored potentials}
%\textcolor{magenta}{The sweet spot of correlated quantum walks: optimal entanglement without transport suppression}

\author{Gaia Forghieri$^{1}$}
\email{gaia.forghieri@unimi.it}
\author{Matteo G. A. Paris$^{1}$}

\affiliation{$^{1}$Dipartimento di Fisica, Università di Milano, I-20133 Milan, Italy}

\date{\today}

\begin{abstract}
Controlling the interplay between particle propagation and quantum correlation generation is a central challenge in quantum transport. Here, we investigate two distinguishable continuous-time quantum walkers evolving on parallel one-dimensional lattices, interacting via distance-dependent potentials. %Starting from an on-site (Hubbard-like) interaction, we show that increasing its strength induces a smooth transition from ballistic, uncorrelated spreading to a tightly bound pair with suppressed transport, while entanglement entropy peaks at an intermediate, optimal value. 
While on-site interactions reproduce the typical bosonic behaviour, extending the interaction to a linear potential over multiple neighbors introduces controlled Bloch-like oscillations and shifts the bound-pair regime to stronger couplings. More generally, we explore a Coulomb-like interaction parameterized by strength, spatial scaling, and decay rate. This reveals a rich \textcolor{black}{dynamical} diagram including four distinct regimes: (i) a high-entropy, oscillatory regime akin to a linear potential; (ii) a strongly localized, bound-pair regime; (iii) a novel intermediate regime combining near-ballistic spreading with strong correlations; and (iv) a weakly interacting, free-propagation regime. Notably, regime (iii) achieves concurrent optimization of transport efficiency and entanglement, offering a sweet spot for correlated quantum dynamics. Our results provide a tool for designing interaction-engineered quantum walks with potential applications in quantum information processing and simulations.
\end{abstract}

\maketitle

\begin{figure*}[t!]
    \centering
    \includegraphics[width=0.95\linewidth]{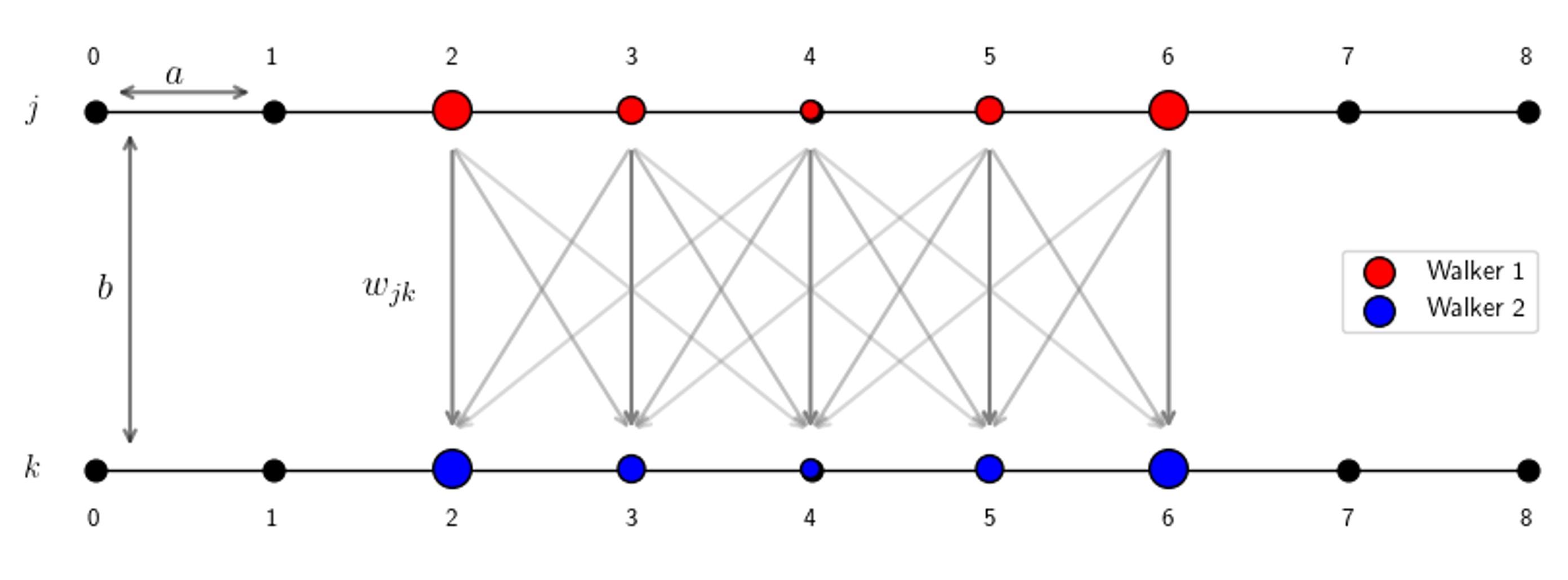}
    \caption{Schematic representation of the system considered in this paper. Two walkers evolve on parallel one-dimensional lattices and interact through a distance-dependent interaction $w_{jk}$ involving multiple neighbors beyond the on-site term. The size of the occupied sites is proportional to the corresponding single-particle occupation probability, and the shade of the interaction arrows is proportional to the strength of the interaction terms. Parameters $a$ and $b$ are respectively the lattice parameter and the interlattice distance.}
    \label{fig:1}
\end{figure*}

\section{Introduction}
Quantum walks represent a powerful tool for modeling coherent transport phenomena and for implementing quantum simulation algorithms \cite{mulken2011,ambainis2003,annoni2024,bottarelli2023,cavazzoni2022,ragazzi2025,candeloro2023,forghieri2026-linkprediction}. In continuous-time quantum walks, a particle evolves under a tight-binding Hamiltonian on a lattice, exhibiting ballistic spreading that contrasts sharply with the diffusive behavior of classical random walks \cite{apers2022}. When two or more quantum walks are considered, their indistinguishability and mutual interactions give rise to a rich variety of correlated transport phenomena, including bound states, entanglement generation, and interaction-induced localization or delocalization \cite{siloi2017,wang2014,preiss2015,yan2019}. These effects are of fundamental interest for understanding non-equilibrium quantum dynamics and can have practical implications for quantum information processing \cite{underwood2012,childs2013,asaka2023,qiang2024,razzoli2024,forghieri2026}, where controlled transport and on-demand entanglement are relevant 
resources.\newline

Previous studies on interacting quantum walks have addressed indistinguishable particles on the same lattice, where exchange statistics play a role, with interactions usually modeled through simple on-site interactions for bosons (Bose-Hubbard model) \cite{lahini2012,winkler2006}, or first-neighbor interactions for fermions \cite{melnikov2016}. This models have revealed interesting dynamical regimes, however with limited flexibility in tailoring the spatial structure of the interaction. In many physically relevant situations, however, the interaction between particles depends explicitly on their distance, as in Coulomb or dipole–dipole potentials. Moreover, when walkers propagate on parallel but distinct lattices, e.g., in coupled waveguides \cite{rai2008,zhou2024,gao2024,raymond25}, trapped-ion arrays \cite{zahringer2010,tamura2020,huerta-alderete2020}, or Rydberg lattices \cite{cote2006,khazali2022,chen2024,palaiodimopoulos2024}, the effective interaction can be engineered via inter-lattice distances and potential shapes. Only limited research has been carried out including interactions beyond the first neighbor \cite{chattaraj2016}, without however analysing in detail the impact of the potential profile. Therefore, these additional degrees of freedom and the way in which they influence transport efficiency and quantum correlations remain largely unexplored.\newline

In this work, we address this gap by systematically investigating the dynamics of two distinguishable continuous-time quantum walks propagating on two parallel one-dimensional lattices. The walkers interact via a distance-dependent potential, which we tune from purely on-site to linear and finally to a generalized Coulomb-like form. Our Hamiltonian includes independent control over the interaction strength, spatial range, and decay profile. Using numerical simulations of the two-particle Schrödinger dynamics, we characterize transport via single-particle properties, such as the single-walker standard deviation and return probability, and quantify correlations via the entanglement entropy and the two-particle meeting probability \cite{stefanak2006}. We also relate our results to the spectral properties of the system, by showing how the interaction profile modifies the two-particle band structure.\newline

Our analysis reveals a rich phenomenology. For purely on-site interactions, we recover the familiar \textcolor{black}{crossover} from ballistic spreading to a bound pair as the interaction strength increases, with entanglement entropy showing a maximum at an intermediate value. Introducing a linear potential leads to (controllable) Bloch-like oscillations without the need of an external drive acting on the system, and shifts the bound regime to stronger interactions. More importantly, using a Coulomb-like potential with tunable strength, range, and decay, yields a two-dimensional \textcolor{black}{dynamical} diagram where four distinct regimes may be identified. Among these, we show a previously unreported regime in which the walkers exhibit near-ballistic spreading (comparable to the free case) while maintaining strong correlations, as measured by both entanglement entropy and meeting probability. This regime has no equivalent considering simpler interaction models, and represents an optimal operating point for correlated quantum transport, where propagation and entanglement are simultaneously enhanced.\newline

The paper is organized as follows. In Sec. \ref{sec:methods} we report the theoretical description of the simulated two-particle system, and describe the quantities used to perform the numerical analysis. Then, in Sec. \ref{sec:results} we report the results of the performed simulations, by first analysing an on-site interaction [Sec. \ref{sec:onsite}], then shifting our focus to a linear potential profile [Sec. \ref{sec:linear}], and finally generalizing our results to a long-range, Coulomb-like potential [Sec. \ref{sec:coulomb}]. We summarize our findings in Sec. \ref{sec:conclusions}, and report additional visual analysis in the Appendix in Sec. \ref{sec:appendix}.

\section{Methods}\label{sec:methods}

\subsection{System definition}

We consider the dynamics of two correlated continuous-time quantum walks propagating on parallel one-dimensional lattices. As each walker propagates on a different lattice, we consider the two particles to be distinguishable. Each lattice consists of $N$ sites, $j\in[0,N-1]$. For each walker, we consider a tight-binding-like Hamiltonian with nearest-neighbor hopping:
\begin{equation}
    H_i = \sum_{j=0}^{N-2} J(|j\rangle\langle j+1| + |j+1\rangle\langle j|)\, , \quad i = 1,2\, ,
\end{equation}
where we set the on-site energy to zero without loss of generality. The hopping amplitude $J$ defines the natural energy scale of the system. Two interacting walkers can then be modeled through a two-particle Hamiltonian of the shape:
\begin{equation}
    H^{(2)} = H_1 \otimes \mathbb{1} + \mathbb{1} \otimes H_2 + H_{\mathrm{int}}\, ,
\end{equation}
where $H_{\mathrm{int}}$ is a two-particle operator which accounts for the interaction between walkers:
\begin{equation}
    H_{\mathrm{int}} = \sum_{j,k=0}^{N-1} w_{jk} |j,k\rangle \langle j,k|\, .
\end{equation}
In our study, we consider a distance-dependent interaction, modeled either through a linear or Coulomb-like potential, as schematically illustrated in Fig. \ref{fig:1}. For this reason, we first define the characteristic lengths of the systems, which are the lattice parameter $a$ (considered equal between lattices 1 and 2) and the distance $b$ between the lattices. The linear potential can be expressed as:
\begin{equation}\label{eq:inter_linear}
    w_{jk}=w_{|j-k|}=\begin{cases}
        J\lambda[1-|j-k|/({\rm n}+1)] & |j-k|\leq{\rm n} \\
        0 & |j-k|>{\rm n}
    \end{cases}\, ,
\end{equation}
where $\lambda$ is a dimensionless quantity representing the relative strength of the interaction with respect to the hopping parameter $J$; and n is the number of nearest neighbors involved in the interaction. The general expression of the Coulomb-like potential is instead:
\begin{align}\label{eq:inter_coulomb}
    w_{jk} = w_{|j-k|} =& \frac{\beta}{\left(\sqrt{b^2 + a^2(j-k)^2}\right)^{\alpha}} \nonumber \\
    =&J\frac{\lambda}{\left(1 + \gamma (j-k)^2\right)^{\alpha/2}}\, ,
\end{align}
where once again we defined the dimensionless quantity $\lambda = \beta/(Jb^\alpha)$ as the relative strength of the interaction. The other two dimensionless quantities are $\gamma = a^2/b^2$ and $\alpha$, which respectively control the spatial scaling and decay of the interaction. Notice that, within this formalism, the true Coulomb potential is recovered for $\alpha=1$. Thus, a reference value of $\lambda$ for $b=1$ and $J=1$ eV, considering interacting charges $q=e$, is:
\begin{align}
    \lambda &= \frac{\beta}{bJ} =\frac{e^2}{4\pi\epsilon_0}\frac{1}{bJ} \approx 14.4\, .
\end{align}
%Notice that the coefficients $w_{jk}$ are only dependent on the difference $j-k$, reflecting translation-invariance along the lattice.
%This parametrization allows us to continuously interpolate between short-range and long-range interactions, providing a unified framework for analysing interaction-driven transport regimes.

Once we have defined the time-indipendent Hamiltonian of the system, the evolution of the walkers can be evaluated through the Schr\"odinger equation:
\begin{equation}
    |\Psi(t)\rangle = e^{-i H^{(2)} \tau/\hbar} |\Psi(0)\rangle\, ,
\end{equation}
where, considering the walkers as non-correlated at the beginning of the evolution, the initial state of the system is a product state:
\begin{equation}\label{eq:psi2}
    |\Psi(0)\rangle = |\psi_1(0)\rangle\otimes|\psi_2(0)\rangle\, .
\end{equation}
In the following, we will refer to time in dimensionless units, by including the hopping parameter $J$ in the Hamiltonian into the definition of $t = J\tau/\hbar$. This is equivalent to a renormalization: $w_{jk}\rightarrow w_{jk}/J$ and $H_i \rightarrow H_i/J$.\newline

\subsection{Numerical analysis}
In the following sections of the paper, we analyze both the effects of the interaction on the spatial evolution of a single walker, and the two-particle quantities characterizing correlation. For both objectives, we analyze the reduced density matrix of the system, by tracing out one of the two walkers from the two-particle density matrix. For example, for walker 1:
\begin{equation}\label{eq:reduced_rho}
    \rho_1(t) = \mathrm{Tr}_2 \{ |\Psi (t)\rangle\langle\Psi(t)| \}\, .
\end{equation}
The single-particle occupation probabilities correspond to the diagonal of this quantity. Thus, we can compute the single-particle standard deviation in time as:
\begin{equation}
    \sigma_1(t) = \sqrt{\sum_{j=0}^{N-1} \rho_{1,jj}(t)\,[j - j_{\mathrm{avg},1}(t)]^2}\, .
\end{equation}
with $j_{\mathrm{avg},1}(t) = \sum_j\rho_{1,jj}(t)j$. The same can be done for the second particle by exchanging $1\leftrightarrow2$. A general expression for $\sigma_i(t)$ is the following:
\begin{equation}\label{eq:sigma}
    \sigma_i(t) = A(t)\, t^{\nu(t)}\, ,
\end{equation}
and ballistic transport is achieved when $\nu=1$ for all $t$ and $A(t)$ is constant, i.e. when $\sigma_i$ increases linearly with time. Then, quantum correlations can be monitored by means of the entanglement entropy of the system:
\begin{equation}
    S = -\sum_l \varrho_{i,l} \ln \varrho_{i,l}\, ,
\end{equation}
where $\varrho_{i,l}$ are the eigenvalues of the reduced density matrix, either of the first or second particle. Another property that is useful to understand correlated transport is the meeting probability \cite{stefanak2006}:
\begin{equation}
    P_{\rm M}(t) = \sum_{j=0}^{N-1} |\langle jj|\Psi(t)\rangle|^2\, ,
\end{equation}
which is the probability for the two walkers to occupy the same site at time $t$, that is, to evolve jointly. Notice that in the case of free transport this quantity becomes, due to the separability of the two-particle state:
\begin{align}
    P_{\rm M}(t) =& \sum_{j=0}^{N-1} \left[|\langle j|\psi_1(t)\rangle|^2|\langle j|\psi_2(t)\rangle|^2\right] \newline \\
    =& \sum_{j=0}^{N-1}P_1(j;t)P_2(j;t)\, ,
\end{align}
where $P_i(j;t)$ is the single-walker occupation probability of site $j$ at time $t$.\newline

\begin{figure*}[t!]
    \centering
    \includegraphics[width=\linewidth]{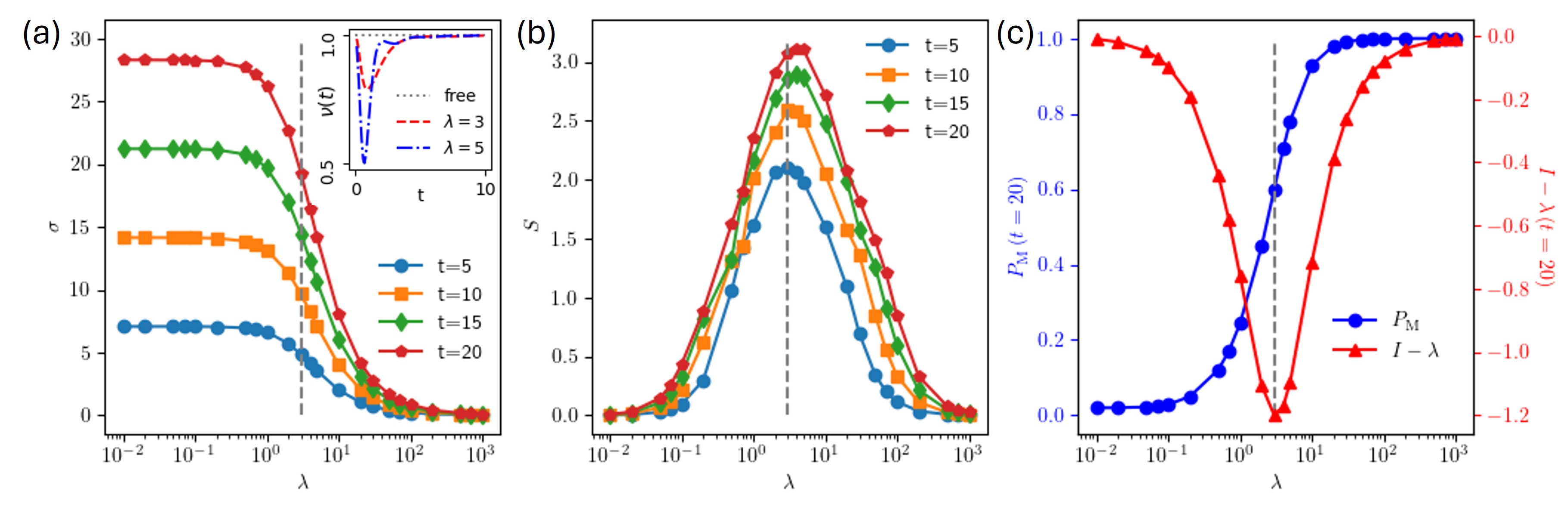}
    \caption{Transport properties of the system in the presence of on-site interaction between walkers. (a) Single-particle standard deviation; the insert shows the exponent of the power law from Eq. \ref{eq:sigma}. (b) Entanglement entropy. (c) Comparison between meeting probability (blue) and expectation value of the interaction operator (red). The grey dashed line is a visual guide for $\lambda=3$.}
    \label{fig:onsite}
\end{figure*}

Finally, we will relate our findings on transport properties with another important aspect of the problem, namely the band structure of the two-particle system. For the sake of this calculation, we consider a periodic extension of the lattice, an approximation that is valid as long as $N\gg\sigma_i(t)$ for every simulated time $t$. Introducing the centre-of-mass and relative coordinates,
\begin{equation}
    M = j + k, \quad m = j - k\, ,
\end{equation}
allows us to use a separable ansatz with the following shape:
\begin{equation}
    \langle M|\Psi\rangle = e^{iqaM/2}|\Phi_q\rangle\, ,
\end{equation}
where $q \in [-\pi/a,\pi/a]$ is the system's wave vector. Introducing this ansatz into the time-indipendent Schr\"odinger equation reduces the description of the two-particle system to an effective one-dimensional Hamiltonian in the relative coordinate:
\begin{equation}\label{eq:ham_rel}
    H_{rel} = \sum_m \left[ 2\cos\!\left(\frac{qa}{2}\right)(|m\rangle\langle m+1| + \text{h.c.}) + w_{|m|} |m\rangle\langle m| \right]\, .
\end{equation}
The two-particle band structure can then be found numerically by solving the eigenvalue problem of the Hamiltonian in Eq. \eqref{eq:ham_rel} for $K=qa\in[-\pi,\pi]$.

\section{Results}\label{sec:results}
In this Section, we show the numerical results of the simulations performed in this work. \textcolor{black}{Unless specified otherwise,} we fixed $N=101$ and initialized each walker at $j_0=k_0=50$. These conditions offer a good approximation of the infinite chain \textcolor{black}{limit up to an evolution time of $t\sim 20$}. All reported time-dependent quantities were obtained with a time resolution of $\delta t = 0.1$.

\subsection{On-site interaction}\label{sec:onsite}
We begin our analysis by considering the case of purely on-site interaction:
\begin{equation}\label{eq:onsite}
    w_{|j-k|} = \lambda\,\delta_{jk}
\end{equation}
which provides a reference framework for understanding the role of more general long-range potentials. This situation is equivalent to setting n=0 in Eq. \eqref{eq:inter_linear}, or alternatively $\alpha\to \infty$ and/or $\gamma\rightarrow\infty$ in Eq. \eqref{eq:inter_coulomb}. In this limit, the interaction affects only configurations in which the two walkers occupy the same lattice site, effectively reproducing a Hubbard-like interaction on parallel chains.\newline

In Fig. \ref{fig:onsite} we show some fundamental transport properties of the walkers' dyamics. First of all, we see from Fig. \ref{fig:onsite}(a) that the standard deviation of the walkers ($\sigma_1(t) = \sigma_2(t) = \sigma(t)$ for symmetry reasons) increases linearly in time for all values of $\lambda$. Indeed, after a small transient time, in all simulations the instantaneous exponent $\nu(t)$ from Eq. \eqref{eq:sigma} approaches unity, indicating that the transport becomes effectively ballistic [see insert in Fig. \ref{fig:onsite}(a)]. On the other hand, the interaction causes a reduction in the prefactor $A(t)$ of the power law in Eq. \eqref{eq:sigma}, effectively reducing the spread of the walkers in time. The \textcolor{black}{crossover} between free transport ($\lambda\to 0$, $A(t)\to A_{\rm free}$) and frozen states ($\lambda \to \infty$, $A(t)\to 0$) is clearly visible in the picture, and is mirrored by the evolution of the band structure as $\lambda$ increases [see Fig. \ref{fig:onsite_regimes} in Appendix \ref{app:onsite}]. Indeed, as it also occurs for bosons \cite{lahini2012,winkler2006}, the presence of the on-site interaction separates a single isolated band \textcolor{black}{(later called ‘bound band')}, to higher energies with respect to the continuum. As the interaction increases, \textcolor{black}{this bound band progressively becomes the dominant contributor for transport [see Fig. \ref{fig:eigenstates_onsite} in Appendix \ref{app:onsite}]. At the same time though, its} slope decreases, and the walkers' velocity drops to zero. This results in a bound pair, with the walkers remaining spatially close to each other, and in turn to the origin, throughout their whole evolution.\newline

We then analyze the generation of entanglement as an effect of the interaction. In Fig. \ref{fig:onsite}(b) we also observe a clear \textcolor{black}{crossover} between weak- and strong-interacting cases, with a maximum value of the entanglement entropy approximately right in the middle of the \textcolor{black}{two regimes} ($\lambda\approx 3$). Indeed, for $\lambda\to 0$, the walkers mostly evolve separately and there is close to no correlation between their states. Thus, the two-particle state can be approximated by a product state:
\begin{equation}
    |\Psi\rangle_{\lambda\to 0}\approx |\psi_1\rangle_{\lambda\to 0} \otimes |\psi_2\rangle_{\lambda\to 0}\, .
\end{equation}
Similarly, when $\lambda\to \infty$, the states of the walkers can be approximated by a $\delta$-function centered at the origin of the two chains, due to their bound-state nature, so that the two-particle state can also be approximated by a product state:
\begin{equation}
    |\Psi\rangle_{\lambda\to\infty}\approx \sum_{j,k}\delta_{j,0}\delta_{k,0} = |\psi_1\rangle_{\lambda\to\infty}\otimes |\psi_2\rangle_{\lambda\to\infty}\, .
\end{equation}
Thus, in both the weak- and strong-interaction limits, the two-particle state remains approximately separable, and the entanglement entropy vanishes. In the intermediate regime, however, the interaction is sufficiently strong to generate non-trivial correlations between the walkers, yet not so strong as to bind them into a localized pair. This balance creates the most favorable conditions for entanglement generation. In turn, this observation provides a practical guideline: maximizing correlation while preserving dynamical transport requires operating at the crossover between the free and bound regimes, where the walkers are correlated but still able to propagate.\newline

The curves in Fig. \ref{fig:onsite}(c) further confirm the behaviors observed in the standard deviation and entanglement entropy. Indeed, the joint probability (blue curve) shows the same \textcolor{black}{crossover} from uncorrelated states for weak interaction ($P_{\rm M}\to 0$) to bound evolution for strong interaction ($P_{\rm M}\to 1$). The curve approximately mirrors the standard deviation, showing a slight asymmetry with respect to the \textcolor{black}{crossover} point at which the entropy is maximum ($\lambda\approx 3$), for which $P_{\rm M}\approx 0.6$ and $\sigma(t)\approx 0.68\,\sigma_{\lambda=0}(t)$. Finally, the red curve in Fig. \ref{fig:onsite}(c) shows the difference between the expectation value of the interaction operator:
\begin{equation}\label{eq:avg_inter}
    I(t) = \sum_{j,k=0}^{N-1} |\langle jk|\Psi(t)\rangle|^2 w_{jk}\, ,
\end{equation}
and the interaction strength $\lambda$. Considering the on-site potential from Eq. \eqref{eq:onsite}, the quantity in Eq. \eqref{eq:avg_inter} becomes:
\begin{equation}
    I(t) = \lambda\sum_{j=0}^{N-1}|\langle jj|\Psi(t)\rangle|^2 = \lambda P_{\rm M}(t)\, .
\end{equation}
Consequently, this quantity tends to $I(t)\rightarrow 0$ for small $\lambda$ and $I(t)\rightarrow \lambda$ for high $\lambda$. In the intermediate regime, $P_{\rm M}<1$ and the average interaction experienced by the walkers will always be lower than $\lambda$. The maximum deviation from $\lambda$ is precisely for $\lambda\approx 3$, thus proving a correspondence with the regime showing maximum entropy.

%{\color{red}By contrast, the deviation from this value is maximized at intermediate coupling, precisely where the system undergoes the crossover between free and bound transport. [ideas | PDF]
%Overall, these results establish a baseline: on-site interactions alone induce a crossover from delocalized to bound transport, accompanied by a transient generation of entanglement and a subsequent suppression of correlations at strong coupling.}

\begin{figure}[t!]
    \centering
    \includegraphics[width=\linewidth]{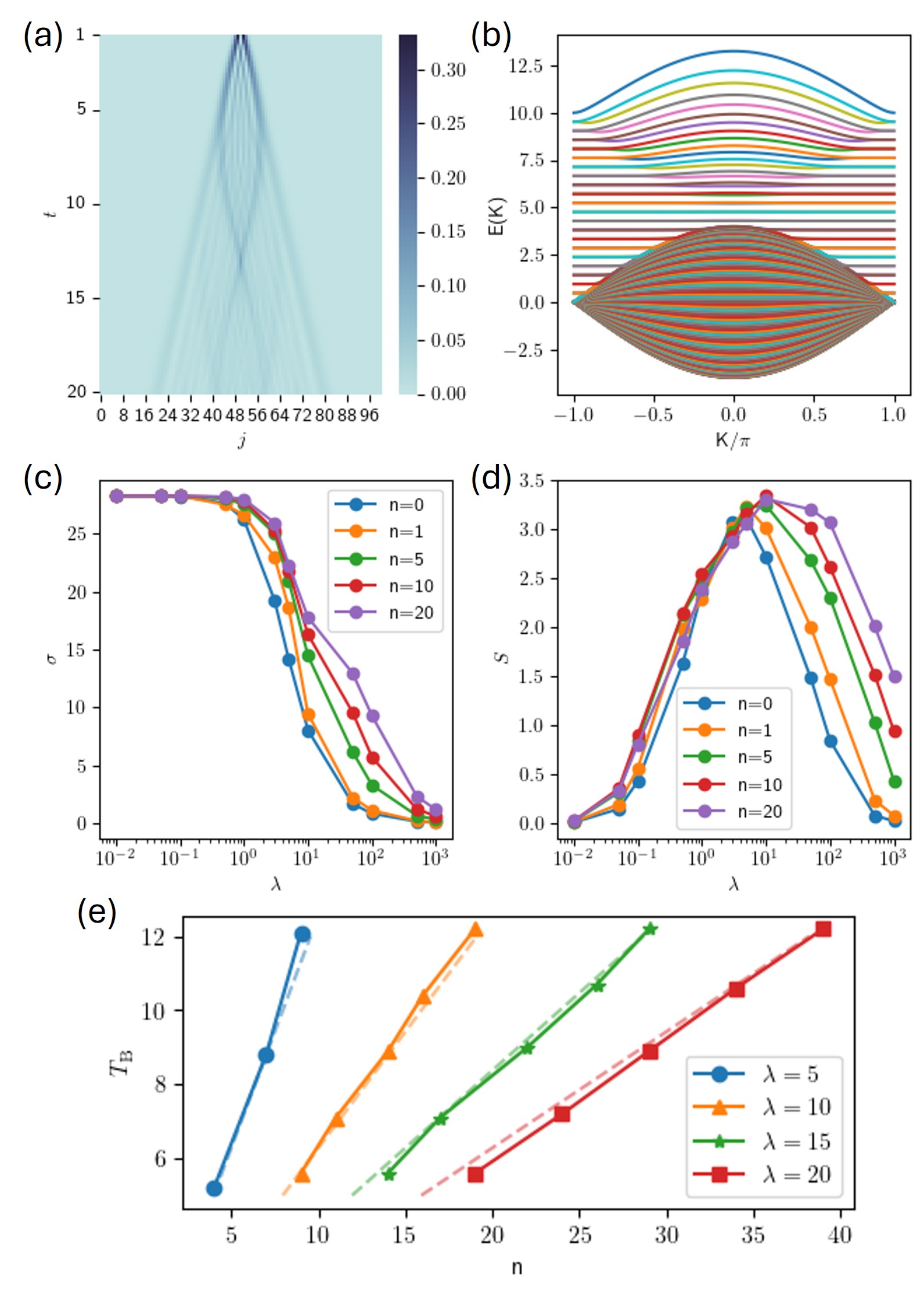}
    \caption{Transport properties in the presence of a linear interaction profile. (a) Single-particle probability distribution and (b) two-particle band structure for parameters $\lambda =10$ and ${\rm n}=20$. (c) Single-particle standard deviation and (d) entanglement entropy, both obtained at time $t=20$ for different numbers of neighbors considered in the interaction. (e) Periods of the Bloch-like oscillations occurring with a linear potential, found numerically through the analysis of the maxima of the return probability in Eq. \eqref{eq:return_prob}.  The dashed lines correspond to the analytical values from the formula in Eq. \eqref{eq:bloch_period}.}
    \label{fig:gradient}
\end{figure}

\subsection{Linear potential}\label{sec:linear}

We now extend our analysis by involving additional neighbors into the interaction. We first consider the most simple case of non-constant interaction, i.e. the linear potential introduced in Eq. \eqref{eq:inter_linear}. This inclusion has the main effect of distributing the interaction over a finite spatial range, thus causing a profound impact on the spectral properties of the system and, consequently, on its transport behavior. We see this in the probability distribution and in the band structure, respectively in Fig. \ref{fig:gradient}(a) and (b). In the latter, due to the non-zero terms of $w_m$ in Eq. \eqref{eq:ham_rel}, a number of bands equal to n is shifted to higher energies. Consequently, they provide a non-negligible contribution to transport. Particularly, the presence of these additional bands causes the \textcolor{black}{bound} band to broaden in energy as n increases [see Fig. \ref{fig:gradient_regimes} in Appendix \ref{app:linear}]. This overall increases its bandwidth, i.e. the walker's velocity, with respect to the on-site case, thus reducing the bound nature of the system for high values of $\lambda$. Indeed, Fig. \ref{fig:gradient}(a) shows that for $\lambda=10$ the single-particle probability distribution is much broader than the one obtained with an on-site interaction of the same strength [compare with Fig. \ref{fig:onsite_regimes}(c) in Appendix \ref{eq:onsite}]. More generally, Fig. \ref{fig:gradient}(c) shows that the standard deviation steadily increases with the inclusion of more and more neighbors, and shifts the bound regime to higher and higher values of $\lambda$. \textcolor{black}{At the same time}, the curves of the entanglement entropy shown in Fig. \ref{fig:gradient}(d) also show this shift, as the entropy for high values of $\lambda$ increases with the number of neighbors. This phenomenon tells us that the system retains correlation for a larger interval of $\lambda$, due to the partial delocalization enabled by the extended interaction profile.\newline

\begin{figure}[t!]
    \centering
    \includegraphics[width=\linewidth]{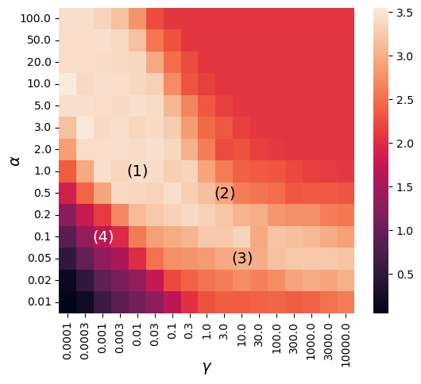}
    \caption{\textcolor{black}{Dynamical diagram showing} the entanglement entropy as a function of parameters $\gamma$ and $\alpha$ defining the scaling of the Coulomb-like potential. Numbers (1) through (4) depict the cases reported in Fig. \ref{fig:regimes}, which are the example cases for the regimes described in the text. All values were obtained for $\lambda=20$ and evolution time $t=20$.}
    \label{fig:entropy}
\end{figure}

That said, another clear effect is visible from the single-particle probability distribution in Fig. \ref{fig:gradient}(b), that is, the presence of an oscillatory behavior mixed with the ballistic propagation of the wavepacket. Indeed, during the time of the evolution, part of the walker returns to its origin, thanks to the presence of backward propagation channels in the band structure \textcolor{black}{[see Appendix \ref{app:linear} for the detailed explanation]}. These oscillations are visually similar to the Bloch oscillations that occur for a single walker in the presence of an external driving force, which is in turn equivalent to a linear energy shift in the on-site energies of the one-dimensional lattice \cite{ahlbrecht2012,preiss2015,chen2024}. In the latter case, the single-particle Bloch period is calculated as:
\begin{equation}\label{eq:bloch_period}
    T_{\rm B} = \frac{2\pi}{\Delta}\, ,
\end{equation}
where $\Delta$ is the energy difference between neighboring sites. In the context of interacting walkers, these Bloch-like oscillations can be understood as a consequence of the effective energy slope felt by each walker due to the interaction. Consequently, in first approximation, we can consider the same formula by setting $\Delta = \lambda/{\rm n}$. Fig. \ref{fig:gradient}(e) shows the comparison between these analytical values and the periods found numerically from the first maximum of the single-particle return probability after $t=0$:
\begin{equation}\label{eq:return_prob}
    P_{\rm R}(t) = \langle j_0|\rho_i(t)|j_0\rangle\, .
\end{equation}

\begin{figure}[b!]
    \centering
    \includegraphics[width=\linewidth]{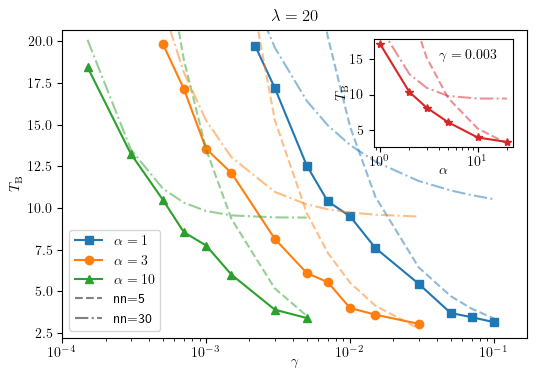}
    \caption{Periods of the Bloch-like oscillations occurring in the regime with low $\gamma$ and high $\alpha$ at fixed $\lambda=20$, obtained numerically through the analysis of the maxima of the return probabiity in Eq. \eqref{eq:return_prob}. The dashed lines correspond to the analytical values obtained through formulas \eqref{eq:bloch_period} and \eqref{eq:delta_eff}, by averaging the interaction step over the first n neighbors.}
    \label{fig:periods}
\end{figure}

\begin{figure*}[ht]
    \centering
    \includegraphics[width=\linewidth]{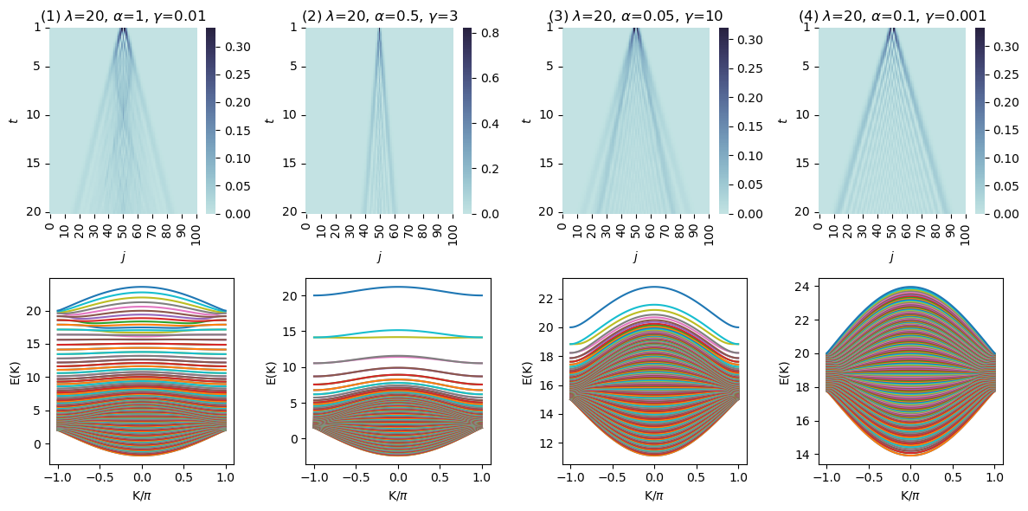}
    \caption{Single-particle probability distribution (above) and band structure (below) for the example trials from Fig. \ref{fig:entropy}, representing regimes (1) through (4) described in the text.}
    \label{fig:regimes}
\end{figure*}

Overall, this analysis shows that the inclusion of interacting neighbors can effectively modify the transport properties of the system through two distinct mechanisms. First, the emergence of multiple bands in the two-particle spectrum broadens the \textcolor{black}{bound} band and increases its slope, which enhances the walkers' velocity compared to the on-site case and shifts the bound regime to higher values of $\lambda$. Second, the approximately linear slope of the interaction over the relevant range of relative distances generates Bloch-like oscillations, with a period $T_{\rm B}\simeq 2 \pi {\rm n} /\lambda$ matching the numerical return probability maxima. These oscillations arise from the effective energy gradient experienced by each walker due to the interaction with the other one, and effectively replace the use of an external drive acting on the system. Together, these findings demonstrate that a linear interaction profile offers a simple yet controllable way to engineer both the average spreading and the coherent oscillatory dynamics of correlated quantum walkers. A natural question arises at this point, concerning how more general, smoothly decaying interactions such as Coulomb-like potentials would affect the dynamics. Indeed, unlike the piecewise linear potential, a Coulomb-like interaction lacks a constant slope and instead exhibits a distance-dependent gradient. As we show in the next Section, this additional flexibility gives rise to a richer \textcolor{black}{dynamical} diagram, where distinct regimes emerge depending on the steepness and spatial range of the decay.

%This is coherent with the fact that increasing the number of interacting neighbours leads to a larger spatial spread and longer oscillation periods (see Fig. \ref{fig:gradient_regimes} in Appendix \ref{}).

%In summary, long-range interactions induce an effective gradient in the two-particle spectrum, thus providing a natural mechanism for the emergence of Bloch oscillations and richer transport phenomenology.

\subsection{Coulomb-like potential}\label{sec:coulomb}

\begin{figure}
    \centering
    \includegraphics[width=0.91\linewidth]{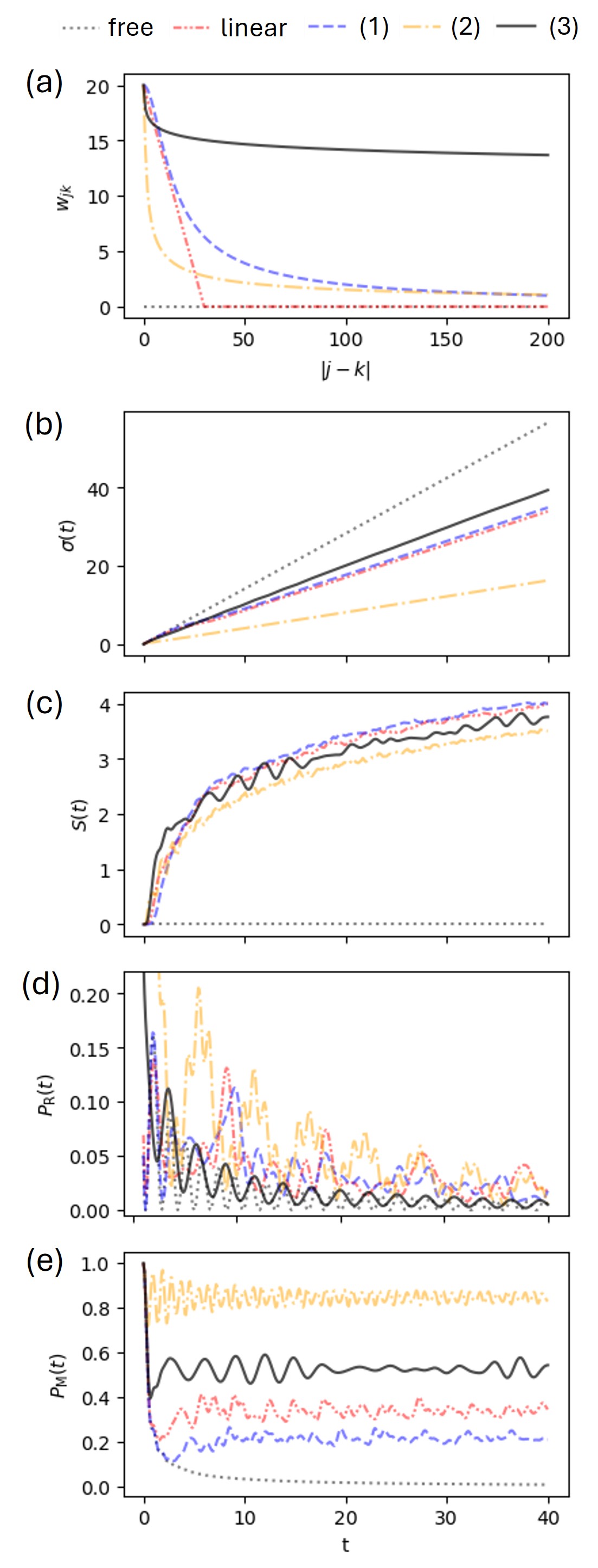}
    \caption{Comparison of the transport properties among regimes (1) through (3) from Figs. \ref{fig:entropy} and \ref{fig:regimes}, the free regime, and the linear potential that most closely resembles the example case (1) ($\lambda=20$, ${\rm n}=30$). For all cases we report: (a) the shapes of the considered potentials; \textcolor{black}{(b) the entanglement entropy; (c)} the single-particle standard deviation; \textcolor{black}{(d)} the return probability; and \textcolor{black}{(e)} the meeting probability.}
    \label{fig:regime_comparison}
\end{figure}

We now generalize our findings to the Coulomb-like case. As in the case of the linear potential, the interaction is distributed among many beighbors. However, the contribution of each neighbor depends on both the scaling parameter $\gamma$ and the decay parameter $\alpha$. For these trials, we fix $\lambda=20$, which is a large enough value to clearly observe the phenomenology of the system at relatively low time ($t_{\rm max}=20$). That said, the same conclusions can be drawn for any other value of $\lambda$, although with slight differences in the absolute values of the entropy in the \textcolor{black}{dynamical} diagram [see Fig. \ref{fig:entropy2} in Appendix \ref{app:coulomb}].\newline

Looking at the behavior of the entropy from Fig. \ref{fig:entropy} as a function of these parameters, our analysis has identified four distinct regimes, exemplified by the four cases (1)-(4) indicated in the picture. We report both the single-walker probability distribution and band structure for all example cases in Fig. \ref{fig:regimes}. Regime (1) corresponds to high values of $\alpha$ (fast decay) and low $\gamma$ (slow scaling), showing high entropy and clearly observable Bloch-like oscillations. \textcolor{black}{Correspondingly}, the band structure closely resembles the one from Fig. \ref{fig:gradient}(a), showing the presence of many bands that are almost equidistant from one another and contribute to transport and to partial back propagation. This is due to the peculiar distribution of the interaction for the first few neighbors, which maintains an approximately constant slope over several lattice sites. Figure \ref{fig:periods} shows the periods obtained numerically from the single-particle probability distribution, which decay fast when increasing either $\gamma$ or $\alpha$. Additionally, it shows a comparison with the periods expected from a linear potential with the same strength $\lambda$ and effective average scaling evaluated as:
\begin{equation}\label{eq:delta_eff}
    \Delta_{\rm eff} = \frac{w_0-w_{\rm n}}{\rm n} = \frac{\lambda}{\rm n}\left[ 1 - \frac{1}{(1\textcolor{black}{+}\gamma\,{\rm n}^2)^{\alpha/2}} \right]\, ,
\end{equation}
for two extreme values of n. For large $\gamma$ or $\alpha$ the decay of the interaction is fast, and the periods are more similar to those obtained with a linear potential spanning only the first $5$ neighbors. Instead, for low $\gamma$ or $\alpha$, the decay of the interaction is much slower, and the situation is comparable to a linear potential spanning $30$ neighbors. That said, when either both parameters are too low or too high, the shape of the potential is too dissimilar from the linear case, and different regimes occur.\newline

Looking back at Fig. \ref{fig:entropy}, regime (2) shows intermediate values of the entropy, asymptotically reaching the values obtained through the on-site potential. Indeed, in this situation, the interaction profile effectively forms a steep well around $|j-k|=0$. \textcolor{black}{Correspondingly}, from the probability distribution and the band structure in Fig. \ref{fig:regimes}, we respectively see the occurrence of a bound pair and of an isolated, flat \textcolor{black}{bound} band at high energies. Equivalently, we can interpret this regime as a continuation of (1), with Bloch-like oscillations occurring at extremely high frequency. As a consequence, the walkers are forced to quickly return to the origin and are unable to propagate.\newline

Regime (4) from Fig. \ref{fig:regimes} is the polar opposite of (2). Indeed, this corresponds to the limit of both $\alpha\rightarrow 0$ and $\gamma\rightarrow 0$, causing the interaction to decay extremely slowly. Overall, this situation asymptotically reproduces a constant interaction potential, whose only effect is the introduction of an additional phase in the evolution of free walkers. Thus, the probability distribution in Fig. \ref{fig:regimes} closely resembles that of the free evolution, and the band structure shows a quasi-continuum of bands forming near the highest-energy band. In addition, the entropy falls to zero because no correlation is formed. Alternatively, we can think of this regime as the limit of regime (1) with extremely large oscillation periods, thus with essentially no back propagation.\newline

Finally, regime (3) corresponds to high values of $\gamma$ and low values of $\alpha$. Thus, the interaction profile is steep for small distances, but decays slowly afterwards. Consequently, the band structure in Fig. \ref{fig:regimes} shows a smooth transition from the main continuum and the highest-energy range of discrete bands. The proximity of the bound band with other bands \textcolor{black}{allows the initial wavefunction to have significant support on multiple propagation channels} and reduces localization, as in regimes (1) and (4), but Bloch-like oscillations are suppressed because the slope of the interaction varies wildly. At the same time, the bound band is separated enough from the continuum to generate sufficient entanglement, as with the on-site interaction. Overall, the value of the entropy is lower but comparable to that in regime (1), as shown in Fig. \ref{fig:entropy}, enhancing at the same time the propagation of the walkers. This indicates an optimal balance between transport efficiency and correlation strength.\newline

In Fig. \ref{fig:regime_comparison} we summarize the transport properties of the regimes described above, compared to the free propagation and linear potential [see Fig. \ref{fig:regime_comparison}(a) for a comparison between the potential profiles considered]. We neglect to show regime (4), which is too similar to the free regime. \textcolor{black}{For these trials, we extended the simulated lattice to $N=201$ sites and fixed the walkers' initial positions to $j_0=k_0=100$. These conditions allow us to simulate the infinite-chain limit up to $t\sim 40$.} Figures \ref{fig:regime_comparison}(b-e) show that, although the free-propagating walker possesses the highest standard deviation over time and the lowest return probability, it also has \textcolor{black}{zero entropy and} the lowest possible meeting probability, due to the complete lack of correlation between walkers. Instead, regime (2), which roughly corresponds to the propagation of a bound pair, shows a meeting probability close to one, implying \textcolor{black}{highly correlated} walkers. However, its standard deviation, \textcolor{black}{although showing a ballistic spread, remains consistently lower than the other trials,} meaning that transport is mostly suppressed. \textcolor{black}{The return probability is also quite high for this case, especially for low times, though it becomes similar to the linear case and regime (1) at larger times.} Regime (1) and (3) \textcolor{black}{overall} work as a middle ground between \textcolor{black}{the free regime and (2)}, with \textcolor{black}{(1) specifically} resembling the transport properties of the linear potential, as previously analyzed. That said, regime (3) stands out as the most favorable for transport engineering, as it combines: (i) a relatively large spreading of the walkers in time, comparable to free transport; (ii) the lowest return probability among all interacting regimes, with almost no increase with respect to free transport; and (iii) a \textcolor{black}{relatively} high meeting probability and entropy, signaling strong correlations throughout the whole evolution. Indeed, although regime (1) possesses a \textcolor{black}{slightly} larger entropy, its meeting probability is much lower, implying that the walkers do not propagate jointly as much as in regime (3). \textcolor{black}{This analysis exemplifies how, even though the entanglement entropy may be similar in magnitude between different trials [see Figs. \ref{fig:entropy} and \ref{fig:regime_comparison}(c)], each case may represent dynamical regimes with widely different transport properties.}\newline

Overall, the results of this section show that the introduction of a Coulomb-like interaction, characterized by the independent tuning of strength, spatial scaling, and decay, leads to a significantly richer phenomenology compared to the simpler interaction models considered in this work and in most of the existing literature. While on-site and linear potentials allow one to control specific features of the dynamics, they inherently constrain the accessible regimes to a limited set of behaviors. In contrast, the combined effect of the three parameters $\lambda$, $\gamma$, and $\alpha$ enables a continuous interpolation between qualitatively different interaction profiles, thereby generating a broader spectrum of transport and correlation properties. This additional flexibility not only enlarges the \textcolor{black}{range} of possible dynamics, but also reveals regimes, such as the intermediate regime (3) discussed above, that cannot be straightforwardly captured within simpler frameworks.\newline

\section{Conclusions}\label{sec:conclusions}
In this work, we have investigated the dynamics of two interacting continuous-time quantum walks propagating on parallel one-dimensional lattices, focusing on how distance-dependent interactions shape transport and correlation properties. Our results show that the nature of the interaction plays a central role in determining the interplay between transport efficiency and correlation generation. %As a reference case, the on-site interaction induces a smooth crossover between free propagation and the formation of a bound pair. In this situation, an intermediate regime emerges in which the interaction is sufficiently strong to generate correlations while still allowing for propagation, resulting in a maximum of the entanglement entropy.
We first analyzed the case of on-site interactions, showing a similar \textcolor{black}{crossover} between free and bound regime to the boson case \cite{lahini2012,winkler2006}, providing a baseline for the subsequent analysis. We then introduced a linear potential profile, which distributes the interaction over multiple neighbors and shifts the bound regime to larger values of $\lambda$. This potential shows the occurrence of Bloch-like oscillations, which establish a first direct connection between the interaction profile and dynamical behaviour, as their period strictly depends on the effective step of the interaction. The fact that these oscillations were obtained without the need of an external drive acting on the system, as was done instead in Refs. \cite{ahlbrecht2012,preiss2015,chen2024}, shows the effectiveness of introducing non-trivial interaction profiles, even as simple as the linear case.

We then studied a more general case in which the potential profile is modeled through a Coulomb-like interaction. This increased flexibility leads to a substantially richer phenomenology, enabling independent control of binding strength, interaction range, and effective energy gradients through the three parameters $\lambda$, $\gamma$ and $\alpha$. As a consequence, it becomes possible to engineer regimes that are not accessible in simpler models. We identified a \textcolor{black}{dynamical} diagram based on the entropy values in the parameter space, where the different regimes correspond to distinct spectral structures and transport behaviors. Specifically, we identified: (1) a linear-like regime with pronounced Bloch-like oscillations and high entropy; (2) a strongly localized regime characterized by suppressed transport; (3) an intermediate regime in which partial delocalization coexists with significant correlations; and (4) a weakly interacting regime approaching free propagation. Among these, regime (3) shows the most promising characteristics for correlated transport. Indeed, the walkers exhibit sustained semi-ballistic spreading with suppressed back-propagation, comparable to the free case, while maintaining a high degree of correlation, as quantified by both the entanglement entropy and the meeting probability. These features identify a non-trivial operating point in which transport and correlation are simultaneously optimized.

Overall, our findings provide a clear guideline for designing interaction-induced quantum transport processes in lattice systems. In particular, the ability to tailor the interaction profile suggests concrete strategies for optimizing correlated propagation, with potential applications in quantum information processing, where both coherent transport and entanglement generation are essential resources.
%Future work may extend this framework in several directions. On the one hand, it would be interesting to explore the role of indistinguishability and quantum statistics, which are expected to further modify the interference and scattering properties of the walkers. On the other hand, implementing these interaction profiles in experimental platforms—such as photonic lattices, cold atoms in optical potentials, or semiconductor nanostructures—would provide a direct route towards realizing interaction-driven quantum transport protocols in controllable settings
%
%By combining numerical simulations with band-structure analysis in the relative coordinate, we have provided a unified description of interaction-driven transport regimes across a wide class of potentials. 

\bibliography{bib.bib}

@article{mulken2011,
title = {Continuous-time quantum walks: Models for coherent transport on complex networks},
journal = {Physics Reports},
volume = {502},
number = {2},
pages = {37-87},
year = {2011},
issn = {0370-1573},
doi = {https://doi.org/10.1016/j.physrep.2011.01.002},
url = {https://www.sciencedirect.com/science/article/pii/S0370157311000184},
author = {Oliver Mülken and Alexander Blumen},
}

@Article{annoni2024,
author={Annoni, Emilio
and Frigerio, Massimo
and Paris, Matteo G. A.},
title={Enhanced quantum transport in chiral quantum walks},
journal={Quantum Information Processing},
year={2024},
month={Mar},
day={23},
volume={23},
number={4},
pages={117},
issn={1573-1332},
doi={10.1007/s11128-024-04331-y},
url={https://doi.org/10.1007/s11128-024-04331-y}
}

@article{bottarelli2023,
    author = {Bottarelli, Alberto and Frigerio, Massimo and Paris, Matteo G. A.},
    title = {Quantum routing of information using chiral quantum walks},
    journal = {AVS Quantum Science},
    volume = {5},
    number = {2},
    pages = {025001},
    year = {2023},
    month = {04},
    issn = {2639-0213},
    doi = {10.1116/5.0146805},
    url = {https://doi.org/10.1116/5.0146805},
}

@article{cavazzoni2022,
    author = {Cavazzoni, Simone and Razzoli, Luca and Bordone, Paolo and Paris, Matteo G. A.},
    title = {Perturbed graphs achieve unit transport efficiency without environmental noise},
    journal = {Physical Review E},
    volume={106},
    number={2},
    pages={024118},
    year={2022},
    publisher={APS},
    doi = {10.1103/PhysRevE.106.024118},
    url = {https://doi.org/10.1103/physreve.106.024118}
}

@Article{ragazzi2025,
AUTHOR = {Ragazzi, Giovanni and Cavazzoni, Simone and Benedetti, Claudia and Bordone, Paolo and Paris, Matteo G. A.},
TITLE = {Scalable Structure for Chiral Quantum Routing},
JOURNAL = {Entropy},
VOLUME = {27},
YEAR = {2025},
NUMBER = {5},
ARTICLE-NUMBER = {498},
URL = {https://www.mdpi.com/1099-4300/27/5/498},
PubMedID = {40422453},
ISSN = {1099-4300},
DOI = {10.3390/e27050498}
}

@article{candeloro2023,
  title={Feedback-Assisted Quantum Search by Continuous-Time Quantum Walks},
  author={Candeloro, Alessandro and Benedetti, Claudia and Genoni, Marco G and Paris, Matteo GA},
  journal={Advanced Quantum Technologies},
  volume={6},
  number={1},
  pages={2200093},
  year={2023},
  publisher={Wiley Online Library},
  url = {https://doi.org/10.1002/qute.202200093},
  doi = {10.1002/qute.202200093}
}

@article{apers2022,
  title = {Quadratic Speedup for Spatial Search by Continuous-Time Quantum Walk},
  author = {Apers, Simon and Chakraborty, Shantanav and Novo, Leonardo and Roland, J\'er\'emie},
  journal = {Phys. Rev. Lett.},
  volume = {129},
  issue = {16},
  pages = {160502},
  numpages = {6},
  year = {2022},
  month = {Oct},
  publisher = {American Physical Society},
  doi = {10.1103/PhysRevLett.129.160502},
  url = {https://link.aps.org/doi/10.1103/PhysRevLett.129.160502}
}

@article{ambainis2003,
author = {Ambainis, Andris},
title = {QUANTUM WALKS AND THEIR ALGORITHMIC APPLICATIONS},
journal = {International Journal of Quantum Information},
volume = {01},
number = {04},
pages = {507-518},
year = {2003},
doi = {10.1142/S0219749903000383},
URL = {https://doi.org/10.1142/S0219749903000383},
eprint = {https://doi.org/10.1142/S0219749903000383}
}

@article{childs2013,
author = {Andrew M. Childs  and David Gosset  and Zak Webb },
title = {Universal Computation by Multiparticle Quantum Walk},
journal = {Science},
volume = {339},
number = {6121},
pages = {791-794},
year = {2013},
doi = {10.1126/science.1229957},
URL = {https://www.science.org/doi/abs/10.1126/science.1229957},
eprint = {https://www.science.org/doi/pdf/10.1126/science.1229957}
}

@article{asaka2023,
  title = {Two-level quantum walkers on directed graphs. I. Universal quantum computing},
  author = {Asaka, Ryo and Sakai, Kazumitsu and Yahagi, Ryoko},
  journal = {Phys. Rev. A},
  volume = {107},
  issue = {2},
  pages = {022415},
  numpages = {14},
  year = {2023},
  month = {Feb},
  publisher = {American Physical Society},
  doi = {10.1103/PhysRevA.107.022415},
  url = {https://link.aps.org/doi/10.1103/PhysRevA.107.022415}
}

@article{qiang2024,
author = {Xiaogang Qiang  and Shixin Ma  and Haijing Song },
title = {Quantum Walk Computing: Theory, Implementation, and Application},
journal = {Intelligent Computing},
volume = {3},
number = {},
pages = {0097},
year = {2024},
doi = {10.34133/icomputing.0097},
URL = {https://spj.science.org/doi/abs/10.34133/icomputing.0097},
eprint = {https://spj.science.org/doi/pdf/10.34133/icomputing.0097},
}

@Article{razzoli2024,
AUTHOR = {Razzoli, Luca and Cenedese, Gabriele and Bondani, Maria and Benenti, Giuliano},
TITLE = {Efficient Implementation of Discrete-Time Quantum Walks on Quantum Computers},
JOURNAL = {Entropy},
VOLUME = {26},
YEAR = {2024},
NUMBER = {4},
ARTICLE-NUMBER = {313},
URL = {https://www.mdpi.com/1099-4300/26/4/313},
PubMedID = {38667867},
ISSN = {1099-4300},
DOI = {10.3390/e26040313}
}

@article{siloi2017,
  title = {Noisy quantum walks of two indistinguishable interacting particles},
  author = {Siloi, Ilaria and Benedetti, Claudia and Piccinini, Enrico and Piilo, Jyrki and Maniscalco, Sabrina and Paris, Matteo G. A. and Bordone, Paolo},
  journal = {Phys. Rev. A},
  volume = {95},
  issue = {2},
  pages = {022106},
  numpages = {8},
  year = {2017},
  month = {Feb},
  publisher = {American Physical Society},
  doi = {10.1103/PhysRevA.95.022106},
  url = {https://link.aps.org/doi/10.1103/PhysRevA.95.022106}
}

@article{preiss2015,
author = {Philipp M. Preiss  and Ruichao Ma  and M. Eric Tai  and Alexander Lukin  and Matthew Rispoli  and Philip Zupancic  and Yoav Lahini  and Rajibul Islam  and Markus Greiner },
title = {Strongly correlated quantum walks in optical lattices},
journal = {Science},
volume = {347},
number = {6227},
pages = {1229-1233},
year = {2015},
doi = {10.1126/science.1260364},
URL = {https://www.science.org/doi/abs/10.1126/science.1260364},
eprint = {https://www.science.org/doi/pdf/10.1126/science.1260364}
}

@article{lahini2012,
  title = {Quantum walk of two interacting bosons},
  author = {Lahini, Yoav and Verbin, Mor and Huber, Sebastian D. and Bromberg, Yaron and Pugatch, Rami and Silberberg, Yaron},
  journal = {Phys. Rev. A},
  volume = {86},
  issue = {1},
  pages = {011603(R)},
  numpages = {5},
  year = {2012},
  month = {Jul},
  publisher = {American Physical Society},
  doi = {10.1103/PhysRevA.86.011603},
  url = {https://link.aps.org/doi/10.1103/PhysRevA.86.011603}
}

@article{ahlbrecht2012,
doi = {10.1088/1367-2630/14/7/073050},
url = {https://doi.org/10.1088/1367-2630/14/7/073050},
year = {2012},
month = {jul},
publisher = {IOP Publishing},
volume = {14},
number = {7},
pages = {073050},
author = {Ahlbrecht, Andre and Alberti, Andrea and Meschede, Dieter and Scholz, Volkher B and Werner, Albert H and Werner, Reinhard F},
title = {Molecular binding in interacting quantum walks},
journal = {New Journal of Physics}
}

@Article{melnikov2016,
author={Melnikov, Alexey A.
and Fedichkin, Leonid E.},
title={Quantum walks of interacting fermions on a cycle graph},
journal={Scientific Reports},
year={2016},
month={Sep},
day={29},
volume={6},
number={1},
pages={34226},
issn={2045-2322},
doi={10.1038/srep34226},
url={https://doi.org/10.1038/srep34226}
}

@article{chen2024,
  title = {Quantum Walks and Correlated Dynamics in an Interacting Synthetic Rydberg Lattice},
  author = {Chen, Tao and Huang, Chenxi and Gadway, Bryce and Covey, Jacob P.},
  journal = {Phys. Rev. Lett.},
  volume = {133},
  issue = {12},
  pages = {120604},
  numpages = {7},
  year = {2024},
  month = {Sep},
  publisher = {American Physical Society},
  doi = {10.1103/PhysRevLett.133.120604},
  url = {https://link.aps.org/doi/10.1103/PhysRevLett.133.120604}
}

@article{wang2014,
  title = {Quantum walks of two interacting anyons in one-dimensional optical lattices},
  author = {Wang, Limin and Wang, Li and Zhang, Yunbo},
  journal = {Phys. Rev. A},
  volume = {90},
  issue = {6},
  pages = {063618},
  numpages = {6},
  year = {2014},
  month = {Dec},
  publisher = {American Physical Society},
  doi = {10.1103/PhysRevA.90.063618},
  url = {https://link.aps.org/doi/10.1103/PhysRevA.90.063618}
}

@article{chattaraj2016,
  title = {Effects of long-range hopping and interactions on quantum walks in ordered and disordered lattices},
  author = {Chattaraj, T. and Krems, R. V.},
  journal = {Phys. Rev. A},
  volume = {94},
  issue = {2},
  pages = {023601},
  numpages = {9},
  year = {2016},
  month = {Aug},
  publisher = {American Physical Society},
  doi = {10.1103/PhysRevA.94.023601},
  url = {https://link.aps.org/doi/10.1103/PhysRevA.94.023601}
}

@article{underwood2012,
  title = {Bose-Hubbard model for universal quantum-walk-based computation},
  author = {Underwood, Michael S. and Feder, David L.},
  journal = {Phys. Rev. A},
  volume = {85},
  issue = {5},
  pages = {052314},
  numpages = {9},
  year = {2012},
  month = {May},
  publisher = {American Physical Society},
  doi = {10.1103/PhysRevA.85.052314},
  url = {https://link.aps.org/doi/10.1103/PhysRevA.85.052314}
}

@article{yan2019,
author = {Zhiguang Yan  and Yu-Ran Zhang  and Ming Gong  and Yulin Wu  and Yarui Zheng  and Shaowei Li  and Can Wang  and Futian Liang  and Jin Lin  and Yu Xu  and Cheng Guo  and Lihua Sun  and Cheng-Zhi Peng  and Keyu Xia  and Hui Deng  and Hao Rong  and J. Q. You  and Franco Nori  and Heng Fan  and Xiaobo Zhu  and Jian-Wei Pan },
title = {Strongly correlated quantum walks with a 12-qubit superconducting processor},
journal = {Science},
volume = {364},
number = {6442},
pages = {753-756},
year = {2019},
doi = {10.1126/science.aaw1611},
URL = {https://www.science.org/doi/abs/10.1126/science.aaw1611},
eprint = {https://www.science.org/doi/pdf/10.1126/science.aaw1611}
}

@Article{zhou2024,
author={Zhou, Wen-Hao
and Wang, Xiao-Wei
and Ren, Ruo-Jing
and Fu, Yu-Xuan
and Chang, Yi-Jun
and Xu, Xiao-Yun
and Tang, Hao
and Jin, Xian-Min},
title={Multi-particle quantum walks on 3D integrated photonic chip},
journal={Light: Science {\&} Applications},
year={2024},
month={Oct},
day={19},
volume={13},
number={1},
pages={296},
issn={2047-7538},
doi={10.1038/s41377-024-01627-7},
url={https://doi.org/10.1038/s41377-024-01627-7}
}

@article{rai2008,
  title = {Transport and quantum walk of nonclassical light in coupled waveguides},
  author = {Rai, Amit and Agarwal, G. S. and Perk, J. H. H.},
  journal = {Phys. Rev. A},
  volume = {78},
  issue = {4},
  pages = {042304},
  numpages = {5},
  year = {2008},
  month = {Oct},
  publisher = {American Physical Society},
  doi = {10.1103/PhysRevA.78.042304},
  url = {https://link.aps.org/doi/10.1103/PhysRevA.78.042304}
}

@article{gao2024,
  title = {Quantum walks of correlated photons in non-Hermitian photonic lattices},
  author = {Gao, Mingyuan and Sheng, Chong and Zhao, Yule and He, Runqiu and Lu, Liangliang and Chen, Wei and Ding, Kun and Zhu, Shining and Liu, Hui},
  journal = {Phys. Rev. B},
  volume = {110},
  issue = {9},
  pages = {094308},
  numpages = {12},
  year = {2024},
  month = {Sep},
  publisher = {American Physical Society},
  doi = {10.1103/PhysRevB.110.094308},
  url = {https://link.aps.org/doi/10.1103/PhysRevB.110.094308}
}

@article{raymond25,
author = {A. Raymond and P. Cathala and M. Morassi and A. Lema\^{i}tre and F. Raineri and S. Ducci and F. Baboux},
journal = {Opt. Express},
number = {22},
pages = {45869--45885},
publisher = {Optica Publishing Group},
title = {Tailoring quantum walks in integrated photonic lattices},
volume = {33},
month = {Nov},
year = {2025},
url = {https://opg.optica.org/oe/abstract.cfm?URI=oe-33-22-45869},
doi = {10.1364/OE.571522},
}

@article{cote2006,
doi = {10.1088/1367-2630/8/8/156},
url = {https://doi.org/10.1088/1367-2630/8/8/156},
year = {2006},
month = {aug},
publisher = {},
volume = {8},
number = {8},
pages = {156},
author = {Côté, Robin and Russell, Alexander and Eyler, Edward E and Gould, Phillip L},
title = {Quantum random walk with Rydberg atoms in an optical lattice},
journal = {New Journal of Physics}
}

@article{palaiodimopoulos2024,
  title = {Chiral quantum router with Rydberg atoms},
  author = {Palaiodimopoulos, Nikolaos E. and Ohler, Simon and Fleischhauer, Michael and Petrosyan, David},
  journal = {Phys. Rev. A},
  volume = {109},
  issue = {3},
  pages = {032622},
  numpages = {11},
  year = {2024},
  month = {Mar},
  publisher = {American Physical Society},
  doi = {10.1103/PhysRevA.109.032622},
  url = {https://link.aps.org/doi/10.1103/PhysRevA.109.032622}
}

@article{khazali2022,
  doi = {10.22331/q-2022-03-03-664},
  url = {https://doi.org/10.22331/q-2022-03-03-664},
  title = {Discrete-Time Quantum-Walk \& Floquet Topological Insulators via Distance-Selective Rydberg-Interaction},
  author = {Khazali, Mohammadsadegh},
  journal = {{Quantum}},
  issn = {2521-327X},
  publisher = {{Verein zur F{\"{o}}rderung des Open Access Publizierens in den Quantenwissenschaften}},
  volume = {6},
  pages = {664},
  month = mar,
  year = {2022}
}

@article{zahringer2010,
  title = {Realization of a Quantum Walk with One and Two Trapped Ions},
  author = {Z\"ahringer, F. and Kirchmair, G. and Gerritsma, R. and Solano, E. and Blatt, R. and Roos, C. F.},
  journal = {Phys. Rev. Lett.},
  volume = {104},
  issue = {10},
  pages = {100503},
  numpages = {4},
  year = {2010},
  month = {Mar},
  publisher = {American Physical Society},
  doi = {10.1103/PhysRevLett.104.100503},
  url = {https://link.aps.org/doi/10.1103/PhysRevLett.104.100503}
}

@article{tamura2020,
  title = {Quantum Walks of a Phonon in Trapped Ions},
  author = {Tamura, Masaya and Mukaiyama, Takashi and Toyoda, Kenji},
  journal = {Phys. Rev. Lett.},
  volume = {124},
  issue = {20},
  pages = {200501},
  numpages = {6},
  year = {2020},
  month = {May},
  publisher = {American Physical Society},
  doi = {10.1103/PhysRevLett.124.200501},
  url = {https://link.aps.org/doi/10.1103/PhysRevLett.124.200501}
}

@Article{huerta-alderete2020,
author={Huerta Alderete, C.
and Singh, Shivani
and Nguyen, Nhung H.
and Zhu, Daiwei
and Balu, Radhakrishnan
and Monroe, Christopher
and Chandrashekar, C. M.
and Linke, Norbert M.},
title={Quantum walks and Dirac cellular automata on a programmable trapped-ion quantum computer},
journal={Nature Communications},
year={2020},
month={Jul},
day={24},
volume={11},
number={1},
pages={3720},
issn={2041-1723},
doi={10.1038/s41467-020-17519-4},
url={https://doi.org/10.1038/s41467-020-17519-4}
}

@article{stefanak2006,
doi = {10.1088/0305-4470/39/48/009},
url = {https://doi.org/10.1088/0305-4470/39/48/009},
year = {2006},
month = {nov},
publisher = {},
volume = {39},
number = {48},
pages = {14965},
author = {Štefanák, M and Kiss, T and Jex, I and Mohring, B},
title = {The meeting problem in the quantum walk},
journal = {Journal of Physics A: Mathematical and General},
}

@misc{forghieri2026,
      title={High-fidelity controlled-phase gates for distinguishable quantum walkers via extended interactions}, 
      author={Gaia Forghieri and Matteo G. A. Paris},
      year={2026},
      eprint={2608.09301},
      archivePrefix={arXiv},
      primaryClass={quant-ph},
      url={https://arxiv.org/abs/2608.09301}, 
}

@Article{winkler2006,
author={Winkler, K.
and Thalhammer, G.
and Lang, F.
and Grimm, R.
and Hecker Denschlag, J.
and Daley, A. J.
and Kantian, A.
and B{\"u}chler, H. P.
and Zoller, P.},
title={Repulsively bound atom pairs in an optical lattice},
journal={Nature},
year={2006},
month={Jun},
day={01},
volume={441}}

@article{forghieri2026-linkprediction,
author = {Forghieri, Gaia and Dubovitskii, Viacheslav and Rossi, Matteo A. C. and Paris, Matteo G. A.},
title = {Link Prediction with Swarms of Chiral Quantum Walks},
journal = {Advanced Quantum Technologies},
volume = {9},
number = {6},
pages = {e70320},
doi = {10.1002/qute.70320},
year = {2026}
}

\begin{widetext}
\newpage
%\begin{multicols}{1}
\section{Appendix}\label{sec:appendix}

In the following, %\textcolor{black}{we first report some considerations on numerical convergence of our simulated system to the infinite-length chain, showing that the parameters used for the trials reported in the main text provide sufficient precision [Fig. \ref{fig:convergence} and Table \ref{tab:convergence}]. Then,}
we report some visually interesting results to exemplify the \textcolor{black}{crossover} between regimes in the presence of an on-site interaction [Fig. \ref{fig:onsite_regimes}] and linear potential [Fig. \ref{fig:gradient_regimes}]. In both pictures and for all cases we report the single-particle probability distribution $\rho_{i,jj}(t)$ (top row), where $\rho_i(t)$ is the reduced density matrix for walker $i$ as evaluated in Eq. \eqref{eq:reduced_rho}. Then, we show the two-particle joint probability at the final time of the evolution, $P_{jk}(t) = |\langle jk|\Psi(t)\rangle|^2$ (central row); and finally, the band structure obtained by numerically solving the eigenvalue problem of the Hamiltonian in Eq. \eqref{eq:ham_rel} (bottom row). \textcolor{black}{For both cases, we also report an analysis of the eigenvectors of the relative-coordinate Hamiltonian from Eq. \eqref{eq:ham_rel} and their contributions to the system's two-particle wavefunction [Figs. \ref{fig:eigenstates_onsite} and \ref{fig:eigenstates_linear}]. This analysis is useful to better explain the processes determining the occurrence of different dynamical regimes. Finallly, we} show the \textcolor{black}{dynamical diagram representing} the entanglement entropy as a function of $\gamma$ and $\alpha$ for different $\lambda$ in the case of the Coulomb-like potential [Fig. \ref{fig:entropy2}], to show the universality of the regimes discussed in Sec. \ref{sec:coulomb}.

%\subsection{\textcolor{black}{Numerical convergence to the infinite chain}}

\subsection{On-site interaction}\label{app:onsite}

\begin{figure}[b!]
    \centering
    \includegraphics[width=0.85\linewidth]{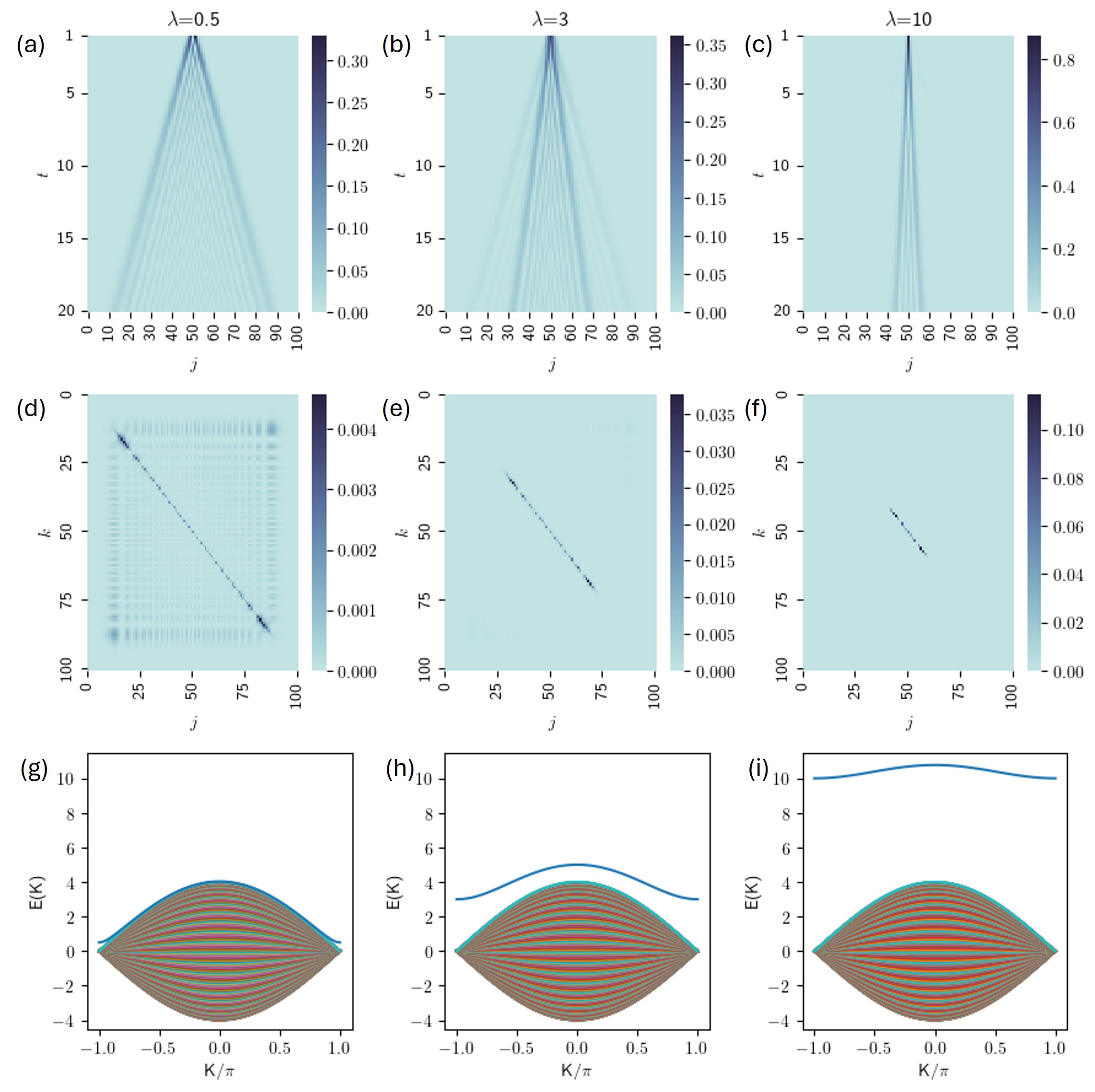}
    \caption{Single-particle probability distribution (top row), joint probability at $t=20$ (central row), and band structure (bottom row) for three cases with on-site interaction of strength $\lambda=0.5$ (left column), $\lambda = 3$ (central column) and $\lambda=10$ (right column).}
    \label{fig:onsite_regimes}
\end{figure}

\begin{figure}
    \centering
    \includegraphics[width=0.85\linewidth]{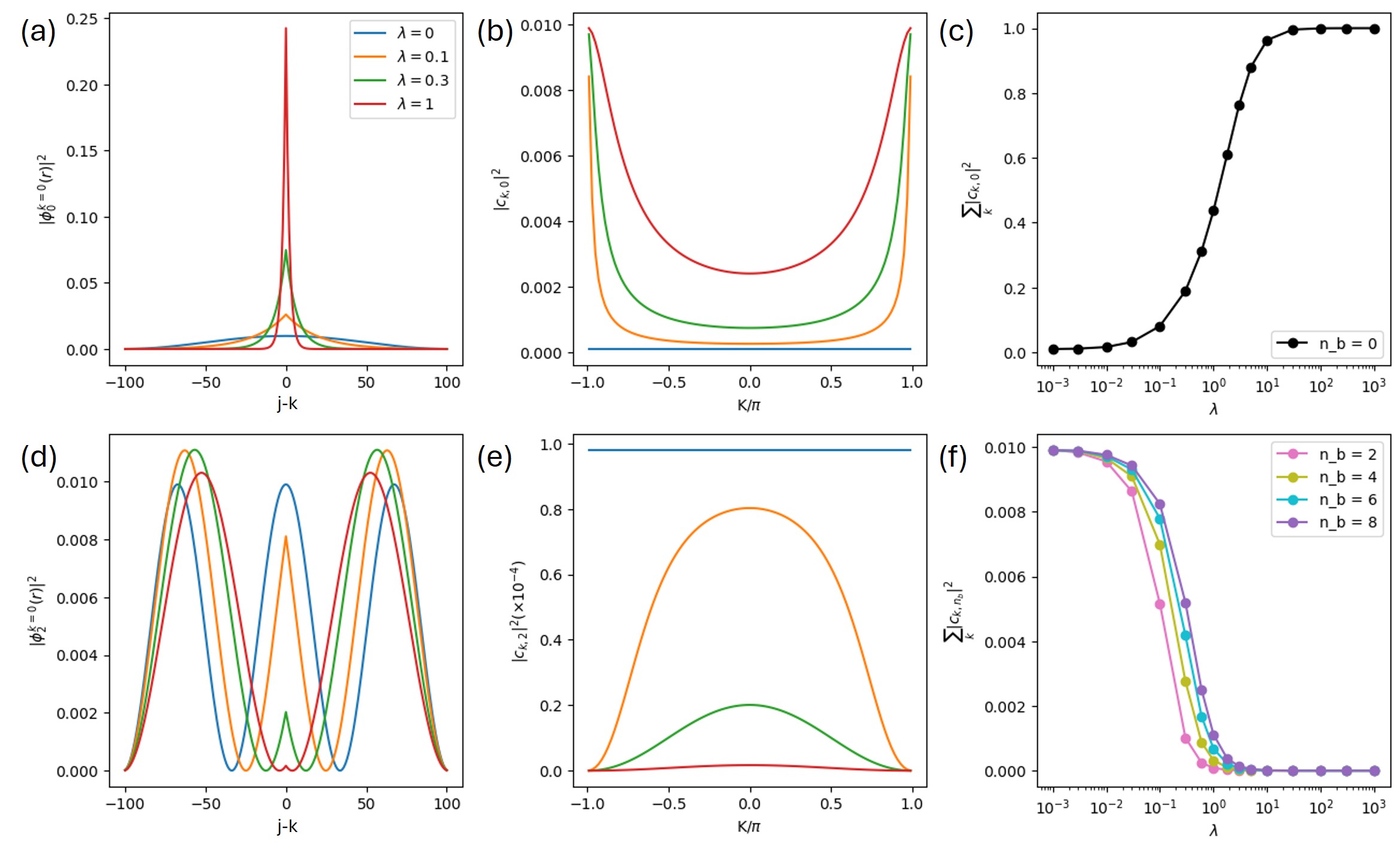}
    \caption{\textcolor{black}{(left) Eigenvectors of the relative-coordinate Hamiltonian for $K=0$ in presence of an on-site interaction (a) for the bound band, $n_b=0$, and (d) for $n_b=2$. (center) $K$-resolved contributions to the system's two-particle wavefunction for (b) $n_b=0$ and (e) $n_b=2$. (right) Contributions of each band summed over $K$ for (c) $n_b=0$ and (f) even $n_b$.}}
    \label{fig:eigenstates_onsite}
\end{figure}

\begin{figure}[b!]
    \centering
    \includegraphics[width=0.85\linewidth]{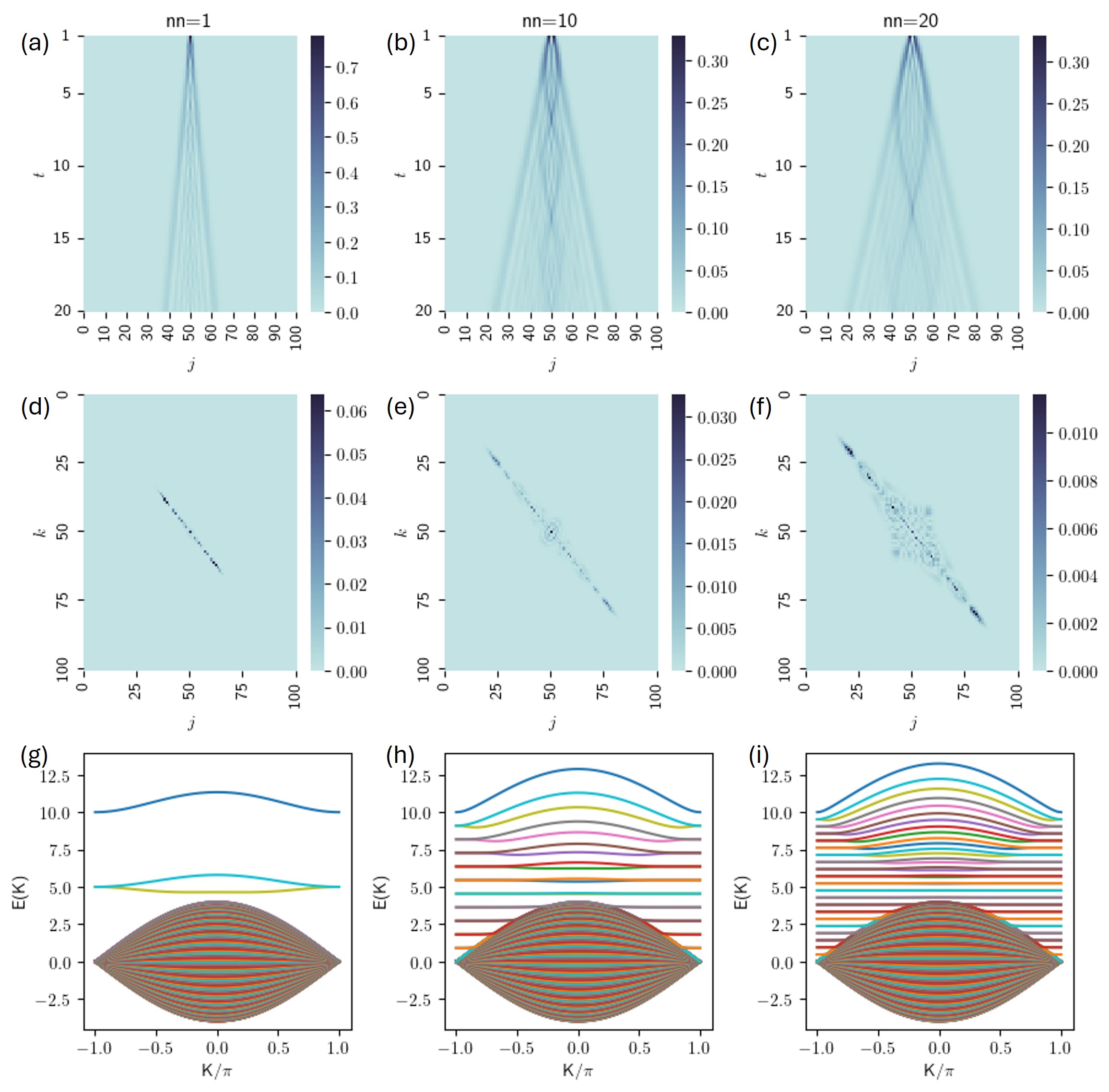}
    \caption{Single-particle probability distribution (top row), joint probability at $t=20$ (central row), and band structure (bottom row) for three cases with linear potential of strength $\lambda=10$ with the inclusion of ${\rm n}=1$ (left column), ${\rm n} = 10$ (central column) and ${\rm n}=20$ (right column) interacting neighbors.}
    \label{fig:gradient_regimes}
\end{figure}

\begin{figure}
    \centering
    \includegraphics[width=0.85\linewidth]{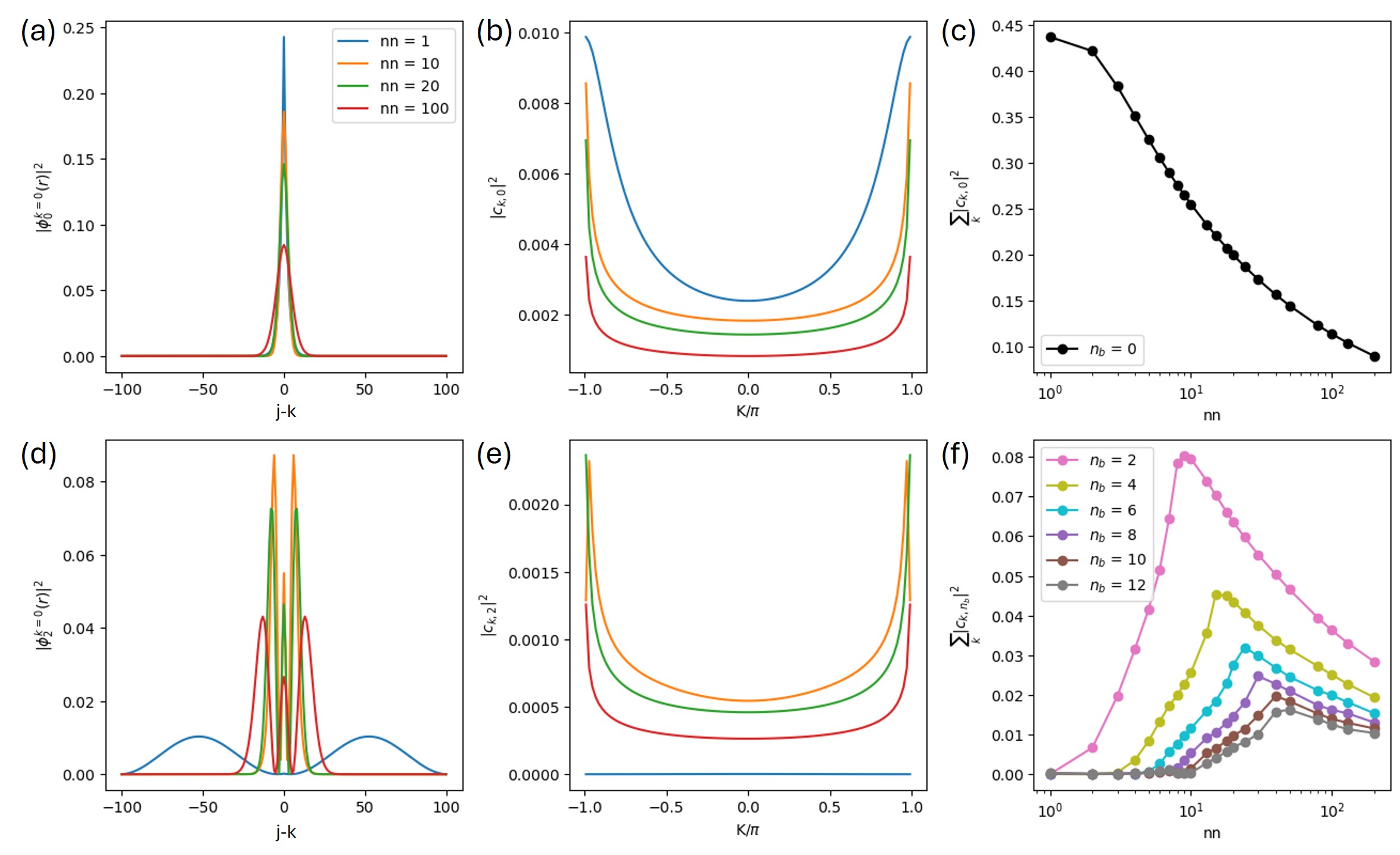}
    \caption{\textcolor{black}{(left) Eigenvectors of the relative-coordinate Hamiltonian for $K=0$ in presence of a linear interaction potential (a) for the bound band, $n_b=0$, and (d) for $n_b=2$. (center) $K$-resolved contributions to the system's two-particle wavefunction for (b) $n_b=0$ and (e) $n_b=2$. (right) Contributions of each band summed over $K$ for (c) $n_b=0$ and (f) even $n_b$.}}
    \label{fig:eigenstates_linear}
\end{figure}

In Fig. \ref{fig:onsite_regimes} we show how the \textcolor{black}{crossover} from free to bound transport with on-site interaction is reflected into the quantities reported in the picture. First, the shape of the single-particle probability distribution confirms that the walkers preserve semi-ballistic transport for all values of $\lambda$, however increasingly reducing their spread. As the interaction increases, the joint probability becomes more and more localized on the diagonal ($j=k$), implying that the walkers propagate jointly and occupy the same sites together. Notice that, visually, the joint probability for $\lambda=3$, corresponding to the \textcolor{black}{intermediate} value between free and bound transport, seems to be completely localized on the diagonal. However, there is a small but non-negligible fraction of the state outside of the diagonal ($j\neq k$), which allows for the generation of entropy, as reported in Fig. \ref{fig:onsite}. As for the band structure, we see that in all cases the spectrum consists of a continuum of dispersive states, corresponding to two non-interacting walkers, with the emergence of a distinct high-energy \textcolor{black}{bound} band as $\lambda$ increases.\newline

\textcolor{black}{In order to explain the link between the emergence of the bound band and localization effects, we show in Fig. \ref{fig:eigenstates_onsite} the characteristics of the system's eigenstates in relative coordinates corresponding to each band, $\phi_{K,n_b}$, where $n_b$ is the band index and $K$ the effective momentum defined in the main text. Keep in mind that, since the system's size is finite, there is only a discrete set of possible $K$-values:}
\begin{equation}\textcolor{black}{
    K = l\frac{2\pi}{N}\, , \quad l = 0,\dots,N-1\, ,}
\end{equation}
\textcolor{black}{which becomes the whole continuous interval $[-\pi,\pi]$ in the limit $N\rightarrow \infty$. These eigenstates can be found from the diagonalization of the relative-coordinate Hamiltonian from Eq. \eqref{eq:ham_rel}. For simplicity, we only show the eigenstates' shape for $K=0$, but an equivalent behavior is also observed for all $K\in[-\pi,\pi]$. Particularly, Fig. \ref{fig:eigenstates_onsite}(a) shows that, as the strength of the on-site interaction increases, the eigenstates for $n_b=0$ (corresponding to the bound band) become increasingly more peaked at $m=j-k=0$, thus resembling more and more the relative-coordinate part of the system's two-particle wavefunction. On the contrary, the eigenstates from the other even-numbered bands [Fig. \ref{fig:eigenstates_onsite}(d) for $n_b=2$] show an odd number of distinct peaks for low $\lambda$, but the middle peak at $m=0$ rapidly decreases as $\lambda$ increases, essentially disappearing for $\lambda>1$. The eigenstates with odd $n_b$ [not shown] show instead an even number of peaks and possess a node at $m=0$. We can now consider these eigenstates to build an orthonormal basis for the two-particle system. First, we need to reconstruct the full two-particle eigenstates:}
\begin{equation}\textcolor{black}{
    \Phi_{K,n_b}(j,k) = \left\langle j,k|\Phi_{K,n_b}\right\rangle = \frac{1}{\sqrt{N}}e^{iKM/2}\phi_{K,n_b}(m)\, ,}
\end{equation}
\textcolor{black}{where we remind that $M=j+k$ is the center-of-mass coordinate. The system's wavefunction can thus be written as a linear combination of these eigenstates:}
\begin{equation}\textcolor{black}{
    \Psi(j,k) = \sum_{K,n_b}c_{K,n_b}\Phi_{K,n_b}(j,k)\, .}
\end{equation}
\textcolor{black}{Consequently, the population of each band at any given $K$ can be retrieved from the following:}
\begin{equation}\textcolor{black}{
    |c_{K,n_b}|^2 = |\left\langle \Phi_{K,n_b}|\Psi \right\rangle|^2 = \left|\sum_{x_1,x_2} \Phi^*_{K,n_b}(j,k)\Psi(j,k)\right|^2\, .}
\end{equation}
\textcolor{black}{Since the two-particle system we consider in this paper is isolated, these contributions don't change in time. Thus, we evaluated them numerically directly from the wavefunction's shape at time $t=0$ [Eq. \eqref{eq:psi2}]. As we can see in Figs. \ref{fig:eigenstates_onsite}(b), the contribution from the bound band progressively increases with $\lambda$ for each $K$. Instead, the contribution from $n_b=2$ (as any other even-numbered band) rapidly decreases to zero [\ref{fig:eigenstates_onsite}(e)], while the contribution from each odd-numbered band is always zero [not shown]. This is in complete agreement with the eigenstates' shapes described previously. Finally, in Figs. \ref{fig:eigenstates_onsite}(c) and (f) we report the sum of the contributions over all values of $K$, confirming that the system's wavefunction occupies the bound band more and more as its gap from the band continuum increases.}\newline

\textcolor{black}{The slope of the bound band,}
\begin{equation}\label{eq:group_vel}
    \textcolor{black}{v_g(k,n_b) = \frac{\partial E_{n_b}(k')}{\partial k'}\bigg|_k\, ,}
\end{equation}
\textcolor{black}{provides a measure of the propagation capability of the corresponding relative-coordinate eigenstates. As Fig. \ref{fig:onsite_regimes}(g-i) show, the bandwidth of the bound} band progressively decreases as the interaction strength grows. \textcolor{black}{At the same time, as we just showed, the wavefunction becomes increasingly dominated by this band, and propagation along different channels is suppressed. Together, these effects indicate a progressive reduction of the mobility of the bound-pair channel, consistent with the decrease of the walkers' dynamical spreading observed in Fig. \ref{fig:onsite}.} This provides a spectral interpretation of the \textcolor{black}{crossover} from free propagation to bound transport described in Sec. \ref{sec:onsite}. Overall, these results closely resemble those for interacting bosons propagating on the same chain \cite{lahini2012,winkler2006}, as the processes occurring in the system are of the same nature.\newline

%Thus, this behavior is directly linked to the formation of localized states. \textcolor{black}{At the same time, as we just showed,} the separation between the bound band and the continuum suppresses \textcolor{black}{the wavefunction's} hybridization between \textcolor{black}{bands, i.e. between different propagation channels, ultimately only occupying the bound band for high $\lambda$. Consequently,} this description explains the reduction of the walkers’ dynamical spreading \textcolor{black}{as $\lambda$ increases. Eventually,} in the strong-coupling limit, the \textcolor{black}{bound} band becomes nearly flat, the walkers' velocities vanish, and a localized bound pair is formed. 

\subsection{Linear potential}\label{app:linear}
In Fig. \ref{fig:gradient_regimes} we show the evolution of the quantities of interest in the presence of a linear potential, and as the number of neighbors involved in the interaction increases. From the single-particle probability distribution we see that the inclusion of additional neighbors causes the presence of Bloch-like oscillations whose period increases with the number of neighbors, or, alternatively, decreases with the step $\Delta$ of the interaction, as described in Sec. \ref{sec:linear}. At the same time, the joint probability changes and, as n increases, starts involving more and more off-diagonal elements. This is related to how, even for large values of the interaction, the entropy is increased, as the two-particle wave-function deviates more and more from a product state. From a spectral point of view, a number of bands equal to double the number of interacting neighbors is shifted to higher energies. \textcolor{black}{Notice that these bands were previously doubly degenerate due to the symmetry $m\leftrightarrow -m$. As these bands move higher up, the energy gap with the bound band decreases. Thus, more and more bands become populated by the system's wavefunction and are available for transport, as can be explained by Fig. \ref{fig:eigenstates_linear}. Indeed, panel (a) shows that the $K=0$ eigenstate of the bound band broadens as the number of interacting neighbors increases. Instead, the peak at $m=0$ starts reappearing in the eigenstates for even $n_b$ [Fig. \ref{fig:eigenstates_linear}(d)], while the eigenstates with odd $n_b$ still show no peak at $m=0$ [not shown]. Because of this behavior, Figs. \ref{fig:eigenstates_linear}(c) and (f) show that the percentage of the system's wavefunction occupying the bound band decreases as the number of interacting neighbors increases, and instead the contributions from all other even bands increases.}\newline

{\textcolor{black}{Going back to the band structures from Fig. \ref{fig:gradient_regimes}(g-i)}, we additionally see that} the two-fold degeneracy of the bands \textcolor{black}{with $n_b\geq 1$} is visibly lifted \textcolor{black}{for those energies closest to the bound band}. \textcolor{black}{Because of this, the bottom band of each pair now possesses for some $K$ an opposite curvature to the bound band's. Particularly, there exists a finite $K$-interval for which the energy gradient of these bands is opposite to that in the bound band. Remember that the group velocity of each component of the system's wavefunction can be evaluated as an energy gradient, as previously seen in Eq. \eqref{eq:group_vel}. Consequently, the participation of these bands to the system's wavefunction introduces the presence of backward-propagation channels, with direction of transport opposite to that in the bound band. The presence of these backward-propagating channels} is the reason for the oscillatory dynamics observed in the probability distribution. That said, as more and more bands approach the \textcolor{black}{bound} band, the bandwidth of the latter increases, thus allowing for higher group velocities of the associated states. This latter effect explains the enhanced spreading of the standard deviation observed in Fig. \ref{fig:gradient} at large $\lambda$. Overall, the band structure reveals that extended interactions enable a further \\textcolor{crossover} between different dynamical regimes. For short interaction ranges, the spectrum is dominated by a single bound band, leading to localization. For longer ranges, the increasing overlap \textcolor{black}{in energy} between multiple bands \textcolor{black}{allows multiple of them, corresponding to different propagation channels, to contribute appreciably to the wavefunction, thus favoring delocalization and} supporting correlated transport over larger distances.\newline

\begin{figure}[b]
    \centering
    \includegraphics[width=\linewidth]{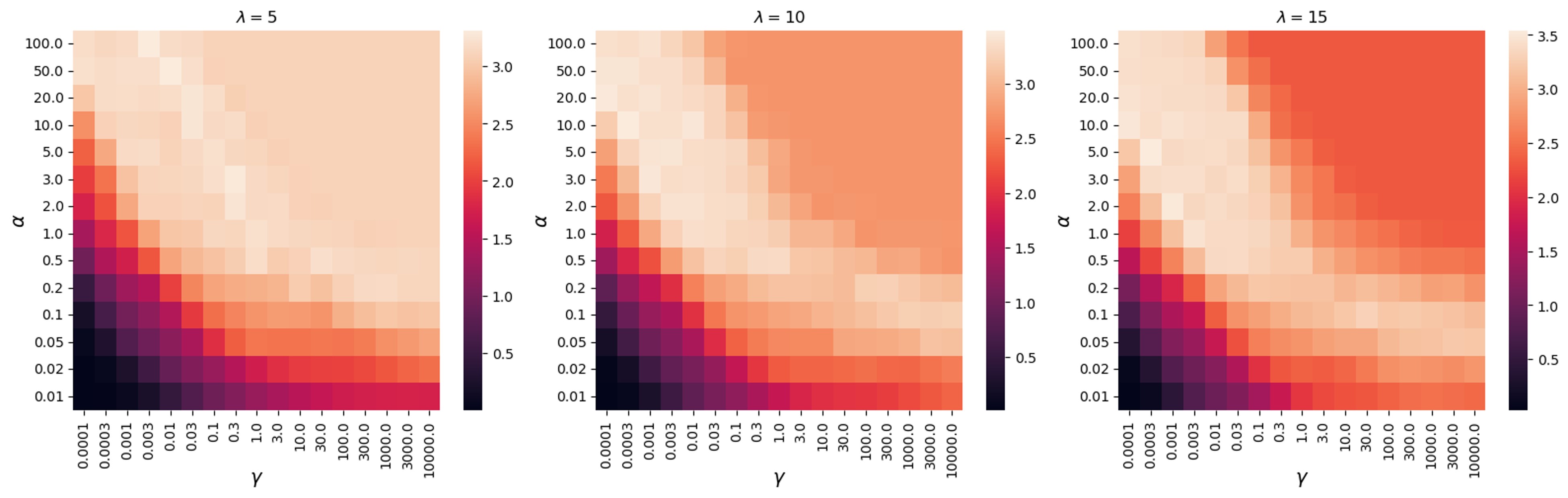}
    \caption{\textcolor{black}{Dynamical diagrams showing} the entanglement entropy as a function of parameters $\gamma$ and $\alpha$ defining the scaling of the Coulomb-like potential, and for different values of the interaction strength $\lambda$. All cases were found by setting $t=20$.}
    \label{fig:entropy2}
\end{figure}

\subsection{\textcolor{black}{Entropy-based dynamical} diagrams with Coulomb-like interaction}\label{app:coulomb}
In Fig. \ref{fig:entropy2} we show the same quantity as in Fig. \ref{fig:entropy} for different values of $\lambda$. One can clearly see that the behavior of the entropy as a function of $\lambda$ and $\alpha$ is equivalent to the one described in Sec. \ref{sec:coulomb}, with the occurrence of the same four regimes detailed in the main text. There are only a few differences with respect to the case with $\lambda=20$. First of all, the absolute values of the entropy slightly increase for regimes (1) and (3) as the strength of the interaction increases. Furthermore, the separation between all regimes is shifted, although very slightly, with respect to the values of the parameters. Lastly, the asymptotical values of the entropy in regime (2), i.e. the quasi-on-site regime, decrease as $\lambda$ increases, coherently with the entropy curves as a function of $\lambda$ in Fig. \ref{sec:onsite}(b).

\end{widetext}
%\end{multicols}
\end{document}